\documentclass[pra,longbibliography,amsmath,amssymb,showkeys,preprint]{revtex4-1}
\usepackage{natbib}
\usepackage{epsfig}
\usepackage{graphicx}% Include figure files
\usepackage{dcolumn}% Align table columns on decimal point
\usepackage{bm}% bold math
\usepackage{setspace}
\usepackage{amsmath}
\usepackage{array}
\usepackage{url}
\usepackage{color}
\usepackage{rotating}

\begin{document}

\title{Wetting dynamics of a collapsing fluid hole}
\author{J.B. Bostwick}
\email[]{jbostwi@clemson.edu}
\homepage[]{http://bostwicklab.sites.clemson.edu}
\affiliation{Department of Mechanical Engineering, Clemson University, Clemson, SC 29631, USA \footnote{Email address for correspondence: jbostwi@clemson.edu}}
\author{J.A. Dijksman}
\email[]{joshua.dijksman@wur.nl}
\affiliation{Department of Physical Chemistry and Soft Matter, Wageningen University, Wageningen, The Netherlands }
\author{M. Shearer}
\email[]{shearer@ncsu.edu}
\affiliation{Department of Mathematics, North Carolina State University, Raleigh, NC 27695, USA}

%\title[Collapse dynamics of a fluid hole in a rotating thin film]{Collapse dynamics of a fluid hole in a rotating thin film}
%\author[J.B. Bostwick and M. Shearer]{J.\ns B.\ns B\ls O\ls S\ls T\ls W\ls I\ls C\ls K$^{1}$\footnote{Email address for correspondence: jbostwi@clemson.edu} \and M.\ns S\ls H\ls E\ls A\ls R\ls E\ls R$^{2}$}
%\affiliation{$^{1}$ Department of Mechanical Engineering, Clemson University, Clemson, SC 29631, USA \\
%$^{2}$ Department of Mathematics, North Carolina State University, Raleigh, NC 27695, USA }
%\date{\today}
%\maketitle

\date{\today}

\begin{abstract}
The collapse dynamics of an axisymmetric fluid cavity that wets the bottom of a rotating bucket bound by vertical sidewalls are studied. Lubrication theory is applied to the governing field equations for the thin film to yield an evolution equation that captures the effect of capillary, gravitational and centrifugal forces on this converging flow. The focus is on the quasi-static spreading regime, whereby contact-line motion is governed by a constitutive law relating the contact-angle to the contact-line speed. The collapse time, as it depends upon the initial hole size, is reported showing that gravity accelerates the collapse process. Surface tension forces dominate the collapse dynamics for small holes leading to a universal power law whose exponent compares favorably to experiments in the literature. Volume dependence is predicted and compared with experiment. Centrifugal forces slow the collapse process and lead to complex dynamics characterized by stalled spreading behavior that separates the large and small hole asymptotic regimes.
\end{abstract}

%\begin{keywords}
%thin films, lubrication theory, contact lines, capillary flows
%\end{keywords}

\maketitle

\section{Introduction}
Coating processes strive to produce uniform thin films on the underlying solid substrate. In certain circumstances a hole can be nucleated in the film. Sometimes these holes disappear and other times they remain as an undesirable defect. As the thickness of the uniform film decreases, it becomes susceptible to instabilities and holes will form in a process called spinodal dewetting \citep{martin2000spinodal}. Controlling the dewetting process allows one to create objects of predetermined size and spatial distribution, as required in many technological applications \citep{gentili2012applications}. For example, micropatterning by dewetting has been used to create desired features in solids ranging from metallic thin films \citep{ferrer2014micro} to soft rubber substrates \citep{martin2001dewetting}. The review by \textcite{geoghegan2003wetting} summarizes the extensive experimental research on wetting/dewetting in polymer films, focusing on the role of pattern formation caused by dewetting. On the scientific side, \textcite{sellier2015estimating} have shown how to estimate the viscosity of a fluid by measuring the collapse time of a nucleated hole.

Hole formation is determined by the stability of the liquid film, which depends on the film thickness $h$ and the sign of the spreading parameter $S\equiv\sigma_{sg}-\left(\sigma_{ls}+\sigma_{lg}\right)$, relating the solid/gas $\sigma_{sg}$, liquid/solid $\sigma_{ls}$ and liquid/gas $\sigma_{lg}$ surface energies. When $S<0$, the film dewets by two mechanisms separated by the scale of the magnitude of the film thickness. In nanometer-sized films ($h<10^{-9}\textrm{m}$), thickness fluctuations lead to intrinsic instabilities that result in spinodal dewetting \citep{martin2000spinodal}. In contrast, mesoscopic films ($10^{-3}\textrm{m} > h > 10^{-8}\textrm{m}$) are neutrally-stable and dewetting occurs by nucleation of a hole via external means, such as capillary suction or air jets \citep{redon1991}. For completely-wetting substrates $S>0$, the nucleated hole is unstable and always collapses. In this paper, we are interested in studying holes in the mesoscopic regime, where surface tension forces play a dominant role in the collapse dynamics.

The experimental literature is filled with novel techniques to nucleate a hole in a thin film. \citet{padday1970cohesive} performed one of the first experimental studies on hole formation in which the critical thickness below which water films ruptured on a variety of surfaces was measured and found to increase with the contact angle. \textcite{taylor1973making} utilized air jets to study hole formation in water and mercury films. They showed that there exists a critical hole size above which larger holes grow and below which smaller holes heal. Experiments by \textcite{redon1991} focus on the rate of hole growth showing that the velocity is independent of film thickness, but critically dependent on the receding contact angle. High velocity drop impact \citep{dhiman2009rupture} and annular retaining dams \citep{diez1992self} have similarly been used. \textcite{backholm2014capillary} have notably studied the interactions between multiple holes in viscous films. Recent experiments by \textcite{mukho09,dijksman2014self} utilize centrifugal forces by rotating an axisymmetric fluid reservoir. These forces drive fluid to the outer edge of the container thereby creating uniform and centered holes. We use an identical geometry in deriving the theoretical model presented here.

With regard to films on partially-wetting substrates, \textcite{sharma1990energetic} showed that two equilibrium holes of different radii are possible for a given contact angle. They used energy arguments (statics) to show that the small and large hole were unstable and stable, respectively, thereby concluding that small enough holes will eventually close. \textcite{moriarty1993dynamic} use lubrication theory to study the dynamics of hole closure for thin films to show that a statically stable hole can be dynamically unstable if there is significant contact-angle hysteresis. \textcite{bankoffdryspot} report dynamic measurements of front velocities, dynamic contact angles and interface shapes, as they depend upon the initial fluid depth. Their results show that the final hole size increases as the initial fluid depth decreases. \textcite{lopez2001stability} conduct a linear stability analysis using a lubrication model with contact line motion to show small holes are unstable to axisymmetric disturbances and large holes eventually become unstable to nonaxisymmetric disturbances. For films in bounded containers, the wetting properties of the sidewalls can also play a significant role in dewetting \citep{lubarda2013shape}.  %The aforementioned studies help to clarify the physics behind wetting/dewetting phenomenon utilized in many micro-and nano-technologies \citep{gentili2012applications}.

Viscous gravity currents occur in industry \citep{ungarish2009book} and in nature \citep{huppert1986intrusion} and can be viewed as a limiting case of the problem we consider here. The flow is primarily horizontal and can be modeled by lubrication theory \citep{huppert1982propagation}. For unbounded flows, the resulting equations admit a self-similar solution of the first kind \citep{gratton1990self} relevant to the dam break problem \citep{ancey2009dam}. The focusing flows that occur in hole collapse can also take a self-similar form of the second kind, although the power law exponent can not be predicted a priori from scaling arguments and must be computed as part of the solution. Experiments on hole collapse can be viewed as convergent viscous gravity currents with \textcite{diez1992self} showing that the size of a hole $a \sim (t_c-t)^{0.762}$, where $t_c$ is the total time for the dry spot to collapse. We recover the exponent predicted by \textcite{diez1992self} in the limit where gravitational forces dominate the collapse dynamics.

In capillary flows, the spreading of a liquid over a solid substrate is controlled by the motion of the contact-line formed at the intersection of the liquid, solid and gas phases. The importance of modeling the contact-line region has been the topic of the reviews by \textcite{dussan79,degennes85,bonn09,snoeijer}, with complex constitutive laws dealing with actual and effective contact angles discussed therein. The most common ad hoc assumption is to allow the fluid to slip at the contact-line in order to relieve the well-known shear stress singularity in the flow field that arises if the no-slip condition is applied \citep{huh71}. Constitutive laws that relate the contact-angle to the contact-line speed, $\theta=f(u_{CL})$, are then introduced in both thin film \citep{greenspan} and irrotational \citep{bostwickS,bostwickARFM} flows. Fluids in unbounded domains, i.e. drops, will spread with characteristic power law in the capillary-dominated limit, as shown in experiments on silicone oil drops by \textcite{tanner79,chen88}. Driving forces such as gravity can alter the spreading exponent \citep{ehrhard91}, while applied thermal fields can cause complex spreading dynamics \citep{bostwick2013thermal}. We use the constitutive law proposed by \textcite{greenspan}, where the contact-angle is linearly related to the contact-line speed, when developing our model for hole collapse. This law is commonly referred to as the Hocking condition \citep{hocking1987damping,bostwickARFM}. Our use of this macroscopic law gives a model that is consistent with the experimental observations in this paper. Characteristic spreading exponents are reported in the i) capillary- and ii) gravity-dominant limits.

Spin coating is a commonly used technique to assist fluids in spreading on solid substrates. One of the first such studies was by \textcite{emslie58}, who analyzed the evolution of an axisymmetric film on a substrate rotating with constant angular velocity to show that initially non-uniform profiles become uniform as a result of centrifugal and viscous forces. If surface tension effects are included in the analysis a capillary ridge may develop near the contact-line of a thin film on a partially-wetting substrate \citep{schwartz,froehlich2009two}. The capillary ridge is seen as a precursor to the fingering instability \textcite[]{melo,fraysse94,spaid96}. \textcite{mckinley2002linear} analyze the linear stability for the equilibrium states of a thin drop on a uniformly rotating substrate, both with and without a central dry patch, and report the growth rate and wavenumber of the critical disturbance. Recent work by \textcite{boettcher2014contact} extend this stability analysis by notably considering general time-dependent base states, from which a critical spreading length from the onset of instability can be inferred. For the hole geometry considered here centrifugal forces retard the collapse dynamics.

We begin by deriving the hydrodynamic field equations that govern the collapse of a fluid cavity. Lubrication theory is utilized to derive an evolution equation for the interface shape. We focus on the quasi-static spreading regime in which the interface shape is static and evolves implicitly through the time-dependent contact-line radius. We report power law forms for the collapse time when i) gravitational or ii) surface tension forces dominate the dynamics. Centrifugal forces that develop in a rotating geometry slow the collapse process and lead to complex dynamics characterized by stalled spreading behavior that separates the large and small hole asymptotic regimes. The role of initial volume is illustrated and compared against experiment. For completeness, the total collapse time is mapped over a large parameter space that depends upon the initial hole size. Lastly, we offer some concluding remarks.

\section{Mathematical formulation}
%\begin{figure}
%\begin{center}
%\includegraphics[width=0.6\textwidth]{defsketch}
%\end{center}
%\caption{\label{fig:defsketch}Definition sketch of the collapsing hole.  }
%\end{figure}
%\begin{figure}
%\begin{center}
%\begin{tabular}{cc}
%($a$) & ($b$) \\
%\includegraphics[width=0.6\textwidth]{defsketch}&
%\includegraphics[width=0.4\textwidth]{defsketchtop2d}
%\end{tabular}
%\end{center}
%\caption{ Equilibria for different Bond number $G$.\label{fig:flowuncoupled}}
%\end{figure}
%\begin{figure}
%\begin{tabular}{l}
% ($a$)  \\
%\includegraphics[width=0.4\textwidth]{defsketch}\\
% ($b$) \\
%\hspace{0.2in}\includegraphics[width=0.3\textwidth]{defsketchtop3d2}
%\end{tabular}
%\caption{Definition sketch of the collapsing hole in ($a$) two-dimensional side view and ($b$) three-dimensional top view.  \label{fig:defsketch}}
%\end{figure}
\begin{figure*}
\begin{center}
\begin{tabular}{ll}
($a$) & ($b$) \\
\includegraphics[width=0.4\textwidth]{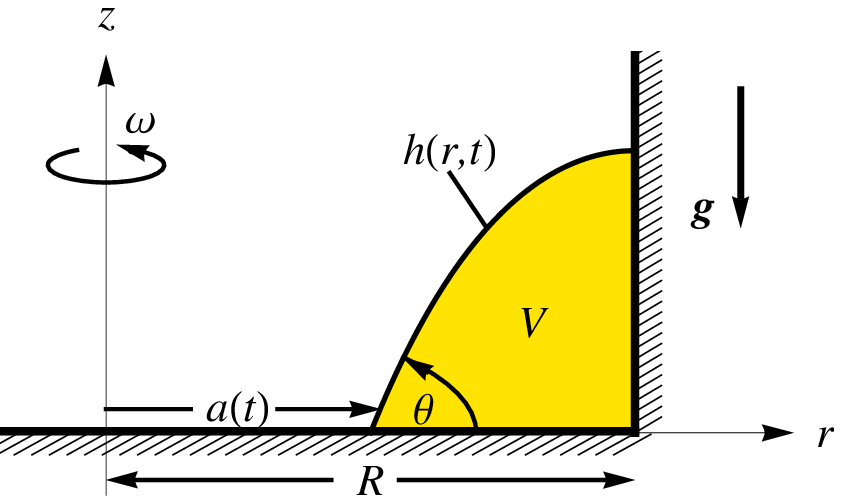}&
\includegraphics[width=0.3\textwidth]{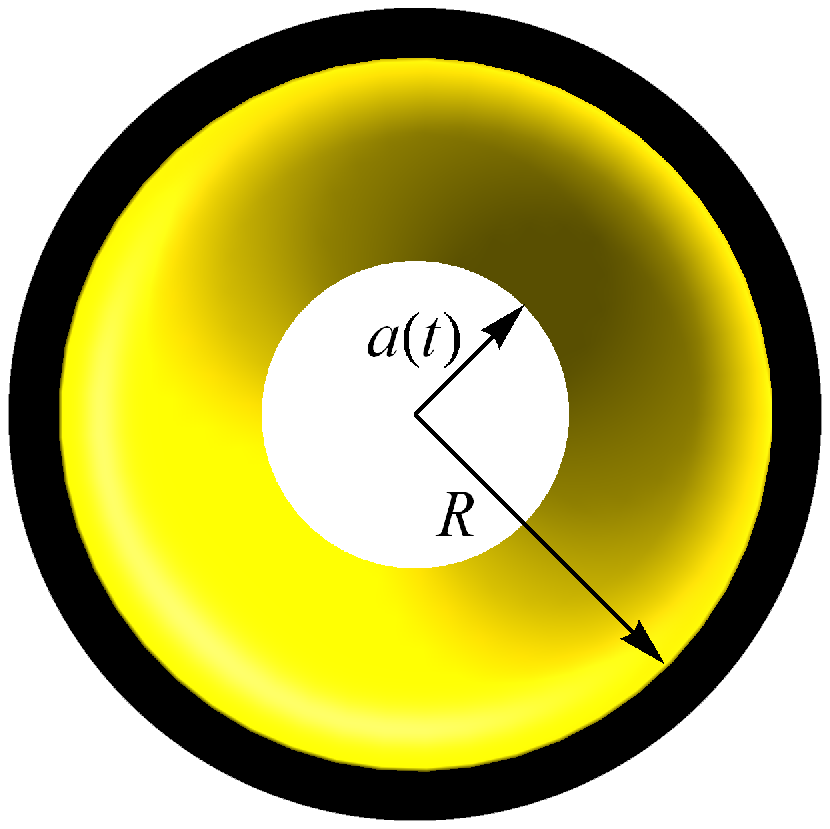}
\end{tabular}
\end{center}
\caption{Definition sketch of the collapsing hole in ($a$) two-dimensional side view and ($b$) three-dimensional top view.  \label{fig:defsketch}}
\end{figure*}
Consider a liquid film wetting the bottom of a solid bucket that is rotating at a constant angular velocity $\omega$ about the vertical axis in axisymmetric cylindrical coordinates $\left(r,z\right)$, as shown in figure~\ref{fig:defsketch}. This incompressible Newtonian fluid has density $\rho$ and dynamic viscosity $\mu$. The liquid and gas phases are separated by an interface $z=h\left(r,t\right)$ ($\partial D$) that is defined on the domain $D$ between the lateral support ($r=R$) and the three-phase moving contact-line $r=a(t)$.

\subsection{Field equations}
The fluid motion is described by the velocity $\boldsymbol{u}=\left(v,w\right)$ and pressure $p$ fields, which satisfy the continuity and Navier-Stokes equations,
%\begin{subequations}\begin{eqnarray} & \boldsymbol{\nabla}\cdot \boldsymbol{u} = 0, \label{incompress} \\ & \rho \left(\frac{\partial \boldsymbol{u}}{\partial t}+ \boldsymbol{u}\cdot \boldsymbol{\nabla}\boldsymbol{u}\right)= \mu \boldsymbol{\nabla}^2 \boldsymbol{u} -\boldsymbol{\nabla}p -\rho g \boldsymbol{\hat{z}} + \rho \omega^2 r \boldsymbol{\hat{r}}, \label{NS}  \end{eqnarray} \end{subequations}
\begin{equation} \begin{split}&\hspace{1in}\boldsymbol{\nabla}\cdot \boldsymbol{u} = 0,  \\ &\rho \left(\frac{\partial \boldsymbol{u}}{\partial t}+ \boldsymbol{u}\cdot \boldsymbol{\nabla}\boldsymbol{u}\right)= \mu \boldsymbol{\nabla}^2 \boldsymbol{u} -\boldsymbol{\nabla}p -\rho g \boldsymbol{\hat{z}} + \rho \omega^2 r \boldsymbol{\hat{r}}. \label{fieldeqn} \end{split}  \end{equation}
Here $g$ is the magnitude of the gravitational acceleration, $\boldsymbol{\hat{r}}=\left(1,0\right)$ the radial unit vector and $\boldsymbol{\hat{z}} = \left(0,1\right)$ the vertical unit vector.

\subsection{Boundary conditions}
The fluid is bounded from below by a rigid substrate $z=0$, where the no-penetration and Navier-slip conditions are enforced, respectively;
\begin{equation}\ w = 0 ,\; v = \beta' \frac{\partial v}{\partial z}. \label{BCsupp}\end{equation}
Here the slip coefficient $\beta'$ is a small number that is introduced to relieve the shear-stress singularity at the contact-line \textcite[]{dussan74}. For reference, alternative methods introduce a precursor layer with disjoining pressure to handle this singularity \textcite[]{popescu2012precursor}. The free surface $z=h\left(r,t\right)$ (liquid/gas interface) bounds the fluid from above and one applies the kinematic condition, balance of normal and shear stresses;
\begin{equation}h_{t} + v h_{r} = w , \quad \boldsymbol{\hat{n}}\cdot \boldsymbol{\mathbf{T}} \cdot \boldsymbol{\hat{n}}= -\sigma \kappa, \quad \boldsymbol{\hat{t}}\cdot \boldsymbol{\mathbf{T}} \cdot \boldsymbol{\hat{n}}= 0. \label{BCfree}\end{equation}
Here $\mathbf{T}$ is the stress tensor and $\sigma$ is the liquid-gas surface tension, while subscripts on the free surface shape $h(r,t)$ denote partial differentiation with respect to the variables $r$ and $t$. The normal $\boldsymbol{\hat{n}}$ and tangent $\boldsymbol{\hat{t}}$ unit vectors are defined with respect to the free surface $h(r,t)$,
\begin{equation} \boldsymbol{\hat{n}} = \left(-h_{r},1\right)/\sqrt{1+h^2_{r}},\;\boldsymbol{\hat{t}} = \left(1,h_{r}\right)/\sqrt{1+h^2_{r}}, \label{surfvecdef}\end{equation}
while the curvature $\kappa$ of that surface is given by
\begin{equation} \kappa = - \frac{\left(r h_{rr} +  h_{r}+ h_r^3\right)}{r\left(1+h_r^2\right)^{3/2}} .\label{meancurv}\end{equation}
We assume neutral wetting conditions on the lateral support ($r=R$),
\begin{equation} h_{r}\big|_{r=R}=0,\; v\big|_{r=R}=0 \label{BCwall}\end{equation}

The contact-line $r=a\left(t\right)$ is located at the intersection of the solid substrate and free surface (cf. figure~\ref{fig:defsketch}). Here
\begin{equation}h(a(t),t)=0, \label{contact}\end{equation}
and the contact-angle $\theta(t)$ is defined by the geometric relationship,
\begin{equation}\frac{\partial h}{\partial r}\left(a(t),t\right)= \tan \theta(t) .\label{CAdef} \end{equation}
At the contact-line, kinematics requires the fluid velocity to equal the contact-line velocity $u_{CL}\equiv v(a(t),t)=\mathrm{d}a/\mathrm{d}t$, which is modeled using a constitutive relationship that relates the contact-line speed to the contact-angle \citep[cf.][]{greenspan,hocking1987damping,hocking1987waves,bostwickS},
\begin{equation} \frac{da}{dt}=\Lambda \left(\theta_A-\theta\right), \label{spreadinglaw}\end{equation}
where $\Lambda > 0$ is an empirical constant and $\theta_A \geq 0$ is the advancing (static) contact-angle. Note that for $\theta>\theta_A$ the fluid displaces gas, $da/dt<0$, in the standard way.  % and $m$ is a spreading exponent---typically $1$ \textcite[]{greenspan} or $3$ \textcite[]{tanner79}.

Finally, we enforce conservation of fluid volume $V_0$,
\begin{equation}2\pi \int_{a(t)}^{R}{r h(r,t)\mathrm{d}r}=V_0. \label{voldef}\end{equation}

\subsection{Lubrication approximation}
The following dimensionless variables are introduced,
%\begin{equation} \tilde{r} = \frac{r}{R}, \; \tilde{z}= \frac{z}{R \theta_0}, \; \tilde{t} = \frac{\sigma}{R \theta_0 \mu} t,\;  \tilde{w}= \frac{\mu}{\sigma}w, \;\tilde{v} = \frac{\mu}{\sigma \theta_0}v, \; \tilde{p}=\frac{R}{\sigma \theta_0} p,\; V=\frac{V_0}{R^3 \theta_0}. \label{scalings}\end{equation}
\begin{equation}\begin{split}&\tilde{r} = \frac{r}{R}, \; \tilde{z}= \frac{z}{R \theta_0}, \; \tilde{t} = \frac{\Lambda \theta_0}{R} t,\;   \tilde{w}= \frac{w}{\Lambda \theta^{2}_0}, \;\tilde{v} = \frac{v}{\Lambda \theta_0 }, \;\\ & \tilde{p}=\frac{R \theta_0 }{\mu \Lambda} p,\; V=\frac{V_0}{R^3 \theta_0}. \label{scalings}\end{split}\end{equation}
Here the size of the lateral support $R$ is used to scale the spatial variables $(r,z)$, the contact-line speed $\Lambda \theta_0$ sets the velocity scale and a viscous pressure scale is used. 

%The following dimensionless groups arise from this choice of scaling;
%\begin{equation}
% C = \frac{\mu \Lambda }{\sigma \theta_0^2} ,\;G^2 = \frac{\rho g R^2}{\sigma} ,\:
% \Omega^2 = \frac{\rho \omega^2 R^3 }{\sigma \theta_0} ,\:
%\beta = \frac{\beta'}{R \theta_0 } ,\end{equation}
%\begin{equation}
% G^2 = \frac{\rho g R^2}{\sigma} ,\:
% \Omega^2 = \frac{\rho \omega^2 R^3 }{\sigma \theta_0} ,\:
%\beta = \frac{\beta'}{R \theta_0 } ,\:\lambda = \frac{\Lambda \theta_0^{2} \mu}{\sigma}\end{equation}
%and slip number $\beta$. %The parameter $\lambda$ is a measure of the contact-line (or wettability) velocity.

The scalings (\ref{scalings}) are applied to the governing equations (\ref{fieldeqn})--(\ref{voldef}) which can then be expanded in terms of the initial contact-angle $\theta_0$, taken to be a small parameter. The leading order expansion (lubrication approximation) gives a reduced set of field equations,
\begin{equation} \frac{1}{r}\left(r v\right)_r + w_z = 0, \; - p_r +v_{zz} + \Omega^2 r =0, \; -C p_z  - G^2=0,\label{fieldS}\end{equation}
where subscripts denote differentiation and the tildes have been dropped for simplicity. Dimensionless constants are given by
\begin{equation}
 C = \frac{\mu \Lambda }{\sigma \theta_0^2} ,\;G^2 = \frac{\rho g R^2}{\sigma} ,\:
 \Omega^2 = \frac{\rho \omega^2 R^3 }{\sigma \theta_0} .\end{equation}
which are the mobility capillary number $C$, Bond number $G^2$, and centrifugal number $\Omega^2$. The boundary conditions on the substrate $z=0$ are given by
\begin{equation} w=0,\;v = \beta v_z,  \label{suppBCS}\end{equation}
with dimensionless slip number $\beta = \beta'/(R \theta_0)$. The reduced free surface boundary conditions on $z=h(r,t)$ are written as
\begin{equation} h_t + v h_r = w ,\; -C p = h_{rr} + \frac{1}{r} h_r.\label{freeBCS}\end{equation}
The dynamic contact-line condition is given by
\begin{equation} \frac{\mathrm{d}a}{\mathrm{d}t}= \left(\theta_A-\theta\right) \label{CLBCs} , \end{equation}
and the volume conservation constraint by
\begin{equation}2\pi \int_{a(t)}^{1}{r h(r,t)\mathrm{d}r}=V. \label{voldefS}\end{equation}

%\begin{eqnarray}  && C = \frac{\mu \kappa \theta_0^{m-3}}{\sigma_0} ,\;\;
% G = \frac{\rho g a_0^2}{\sigma_0} ,\;\;
% \Omega^2 = \frac{\rho \omega^2 a_0^3 \theta_0^{2-m}}{\mu \kappa} ,\nonumber \\
%&& \beta = \frac{\beta'}{a_0 \theta_0 } ,  \;\;
% \Delta C = \frac{\mu \kappa \theta_0^{m-1} }{\gamma\left(T_0 - T_{\infty}\right)} ,\;\;
% B = \frac{h_g a_0 \theta_0 }{k} ,\end{eqnarray}
%
\subsection{Derivation of evolution equation}
We begin by constructing a solution to the governing equations (\ref{fieldS})--(\ref{freeBCS}) that depends implicitly on the free surface shape $h$. The pressure is computed from the vertical component of the Navier-Stokes equation (\ref{fieldS}) and normal stress balance on the free surface (\ref{freeBCS}),
\begin{equation}C p = G^2 \left(h-z\right)  -\left(h_{rr} + \frac{1}{r} h_r\right). \label{presssol}\end{equation}
The radial velocity field is calculated from the radial component of the Navier-Stokes equations (\ref{fieldS}), Navier-slip condition (\ref{suppBCS}) and tangential stress balance (\ref{freeBCS}),
\begin{equation} v = \left(p_r - \Omega^2 r\right)\left(\frac{1}{2} z^2 - \left(z+\beta\right)h \right). \label{ursol}\end{equation}
We then use the reduced continuity equation (\ref{fieldS}) and no-penetration condition (\ref{suppBCS}) to compute the vertical velocity
\begin{equation} w = - \left(p_{rr} + \frac{1}{r} p_r- 2\Omega^2\right) \left(\frac{1}{6}z^3 - h\left(\frac{1}{2}z^2 + \beta z\right)\right). \label{uzsol}\end{equation}

Finally, we apply the fields defined in (\ref{presssol})--(\ref{uzsol}) to the depth-averaged continuity equation $h_t + \left(1/r\right) \left(r q\right)_r=0$, with $q$ the net radial flux, to generate the evolution equation,
\begin{widetext}
\begin{equation} C h_t + \frac{1}{r}\left(r\left(\left(h_{rr} + \frac{1}{r} h_r-G^2h\right)_r+ \Omega^2 r\right)\left(\frac{1}{3}h^3 + \beta h^2\right)  \right)_r = 0 .\label{evoleq}\end{equation}
\end{widetext}
The motion of the fluid interface is governed by the evolution equation (\ref{evoleq}), the dimensionless form of the contact-line conditions (\ref{contact})--(\ref{spreadinglaw}) and conservation of volume constraint (\ref{voldef}). Once the free surface shape $h$ is known, pressure $p$ and velocity $(v,w)$ fields are then computed from (\ref{presssol})--(\ref{uzsol}).

\subsection{Quasi-static spreading ($C\rightarrow 0$)}
In this paper, we focus on the quasi-static limit $C\rightarrow 0$ proposed by \textcite{greenspan} that has been utilized by a number of authors \citep[e.g.][]{rosenblat,ehrhard91,smith95}. The approximation is justified by noting that typical spreading rates can be on the order of microns per second, which is much slower than the velocity scale obtained by balancing viscosity with surface tension. Quasi-static spreading describes a static droplet shape that is parameterized by the contact-line radius $a$, which evolves according to the unsteady dynamic contact-line condition (\ref{CLBCs}). More precisely, the free surface shape evolves implicitly through the time-dependent contact-line radius.  The leading order problem consists of a steady droplet shape with no contact-line motion, therefore we may set the slip number $\beta=0$.

The steady evolution equation (\ref{evoleq}) is integrated to yield an equation governing the steady droplet shape,
\begin{equation} \left(h_{rr} + \frac{1}{r} h_{r} - G^2 h\right)_{r} + \Omega^2 r  = 0,\quad r\in [a,1]. \label{quasistaticeq}\end{equation}
where the integration constant is set to zero to enforce the no-flux condition on the bounding surface (\ref{BCwall}). The dynamic contact-line condition,
\begin{equation} \frac{\mathrm{d}a}{\mathrm{d}t}  =  \left(\theta_A-h_r(a)\right), \label{spreadlawS}\end{equation}
then governs the rate of spreading.

\section{Results}
In this section we describe the dynamics of hole collapse by reporting interface shapes and the time-evolution of the contact-line radius (equivalently, hole size). Each hole we consider here eventually closes; i.e. there are no equilibria with finite hole size. Hence, one important metric is the time required for the hole to completely collapse $t_c$ into a film. We compute $t_c$ by integrating (\ref{spreadlawS}) with initial conditions $a(0)=a_0$ until $a(t_c)=0$,
\begin{equation} \int_0^{t_c}{\mathrm{d}t}=\int_{a_0}^{0}{\frac{\mathrm{d}a}{\theta_A-h_r(a)}}. \label{collapsetime} \end{equation}
Herein, we consider the case of completely-wetting substrate $\theta_A=0$ although it would be straightforward to consider the more general case $\theta_A\neq0$. We begin with the capillary-dominated regime as a base case and then focus on how gravitational $G$ and centrifugal $\Omega^2$ forces affect the collapse dynamics.

\subsection{Capillary-dominated collapse}
\begin{figure}
\begin{center}
\includegraphics[width=0.4\textwidth]{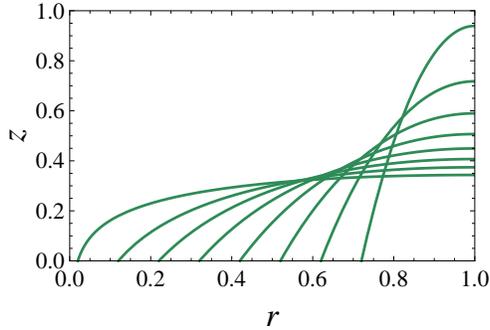}
\end{center}
\caption{ Capillary-dominated collapse ($G=0,\Omega^2=0$): equilibrium interface shapes, as they depend upon the contact-line radius $a$ for $V=1$. \label{fig:shapesG0}}
\end{figure}

\begin{figure}
\begin{center}
\includegraphics[width=0.4\textwidth]{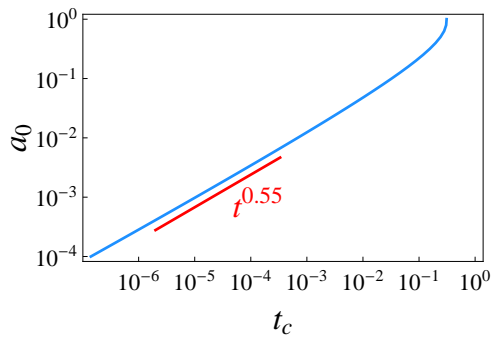}
\end{center}
\caption{Capillary-dominated collapse ($G=0, \Omega^2=0$): initial contact-line radius $a_0$ against collapse time $t_c$ exhibits power law behavior $a_0\sim t^{0.55}$ as $a_0\rightarrow 0$ for $V=1$.  \label{fig:collapseScaleC}}
\end{figure}

The solution of the steady evolution equation (\ref{evoleq}) when surface tension forces dominate the collapse dynamics, $G=0,\Omega^2=0$, is given by
\begin{equation}h(r)=\left(\frac{2V}{\pi}\right)\frac{r^2-a^2+2\ln(a/r)}{a^4-4a^2+3+4\ln(a)}.\label{solG0}\end{equation}
Figure~\ref{fig:shapesG0} plots the corresponding interface shapes as they depend upon the contact-line radius $a$ to show the evolution during the collapse process. The collapse time is obtained from (\ref{collapsetime},\ref{solG0}) to yield
\begin{widetext}
\begin{equation}t_c=\frac{\pi}{48V}\left(4\pi^2+3a_0^2\left(a_0^2-6\right)+ 24\left(\ln(a_0)\ln(1+a_0)+\mathrm{Li}_2(-a_0)-\mathrm{Li}_2(1-a_0)\right)\right),\label{tcG0}\end{equation}
\end{widetext}
where $\mathrm{Li}_2$ is the dilogarithm function. Figure~\ref{fig:collapseScaleC} plots the initial radius $a_0$ against the collapse time $t_c$. In the asymptotic small hole limit $a_0\rightarrow 0$, the capillary-dominant collapse time (\ref{tcG0}) takes the functional form
\begin{equation} t_c \sim \frac{\pi}{8V}\left(4\ln(a_0^{-1})-1\right)a_0^2, \quad \textrm{as} \: a_0 \rightarrow 0\label{collapsesmall}\end{equation}
with a lower bound given by $a_0\sim t^{0.5}$.


\begin{figure}
\begin{center}
\includegraphics[width=0.4\textwidth]{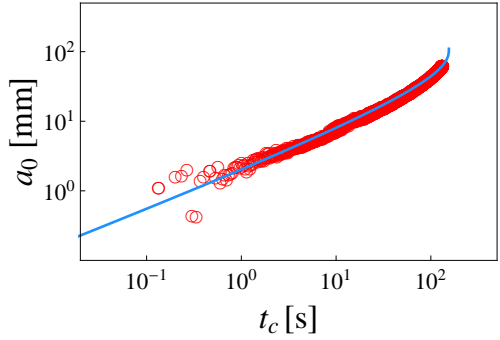}
\end{center}
\caption{Dimensional contact-line radius $a_0$[mm] against collapse time $t_c$[s] for capillary-dominated collapse ($G=0, \Omega^2=0$) with $\mu=10$~mPa$\cdot$s, $\sigma=0.02$~N/m, $R=6.5$~cm and $V_0=38.49$~$\textrm{cm}^3$ with fitted parameter $\Lambda=0.319$~mm/s. Symbols are experimental data points from \textcite[Fig. 2b]{dijksman2014self}. \label{fig:collapseScaleCexp}}
\end{figure}

Unlike droplet spreading \citep{tanner79,chen88}, the asymptotic form (\ref{collapsesmall}) does not admit a specific power law because of the logarithmic term. Although, as shown in Figure~\ref{fig:collapseScaleC}, the collapse dynamics follows the power law $a_0\sim t^{0.55}$ over a range of $a_0$. This particular exponent has recently been reported in experiments on hole collapse by \textcite[Fig. 2b]{dijksman2014self}. Figure~\ref{fig:collapseScaleCexp} shows that our prediction for the collapse time (\ref{tcG0}) compares favorably to these experiments over a range that encompasses the logarithmic correction and is not defined by a single exponent. By fitting our theoretical prediction to the experimental data, we can obtain an estimate for the empirical constant $\Lambda=0.319$~mm/s in Eq.~(\ref{spreadinglaw}).

For the special case $a_0=1$, where the film initially completely wets the bucket sidewall, the collapse time from (\ref{tcG0}) is given by
\begin{equation}t_c=\frac{\pi}{48V}\left(2\pi^2-15\right).\label{wallcollapseG0}\end{equation}
Note that (\ref{wallcollapseG0}) is an upper bound on the total collapse time.

\subsection{Gravity-dominated collapse}

\begin{figure}
\begin{center}
\includegraphics[width=0.4\textwidth]{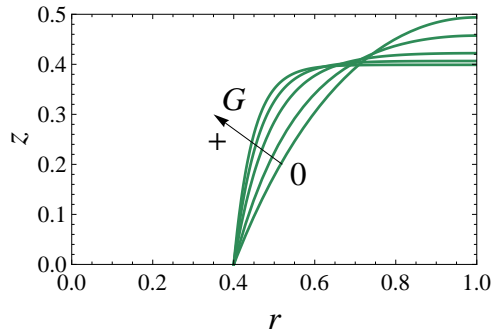}
\end{center}
\caption{ Interface shapes for fixed contact-line radius $a=0.4$, as it depends upon Bond number $G$, shows that gravitational forces tend to flatten the film while increasing the contact-angle.  \label{fig:equilG}}
\end{figure}

When gravitational forces $G\neq0$ are included in the model with $\Omega^2=0$, the solution of (\ref{quasistaticeq}) is given by
\begin{widetext}
\begin{equation}h(r)=\left(\frac{GV}{\pi}\right)\frac{\mathcal{I}_1(G)\left(I_0(Gr)-I_0(Ga)\right)+\left(K_0(Gr)-K_0(Ga)\right)}{\mathcal{I}_1(G)\left(\left(a^2-1\right)K_0(Ga)+2aK_1(Ga)\right)+\left(\left(a^2-1\right)I_0(Ga)-2aI_1(Ga)\right)},\label{solG}
\end{equation}
\end{widetext}
where $I_n,K_n$ are the modified Bessel functions of order $n$ and we have defined $\mathcal{I}_1(G)\equiv I_1(G)/K_1(G)$. Gravity tends to flatten the interface and increase the contact-angle, as shown in Figure~\ref{fig:equilG}.

\begin{figure}
\begin{center}
\includegraphics[width=0.4\textwidth]{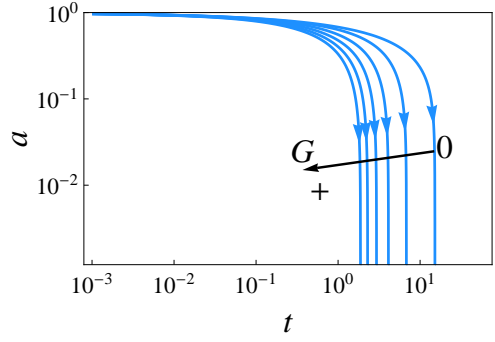}
\end{center}
\caption{ Evolution of the contact-line radius $a$ against time $t$ for initial conditions $a_0=0.99$ and $V=1$, $\Omega^2=0$, as it depends upon Bond number $G$, shows that gravitational forces promote collapse. \label{fig:timeG}}
\end{figure}

\begin{figure}
\begin{center}
\includegraphics[width=0.4\textwidth]{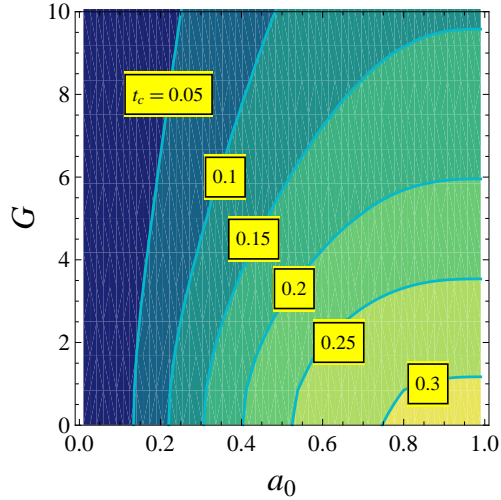}
\end{center}
\caption{ Collapse time $t_c$ against $G$ and initial contact-line radius $a_0$ for $V=1$,$\Omega^2=0$. \label{fig:collapseSessile}}
\end{figure}

\begin{figure}
\begin{center}
\includegraphics[width=0.4\textwidth]{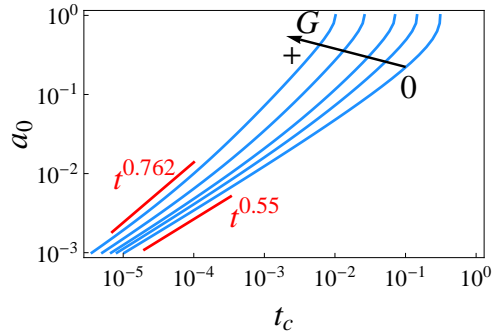}
\end{center}
\caption{Initial contact-line radius $a_0$ against collapse time $t_c$ is bound by asymptotic power law behavior $a_0\sim t^{0.55}$ for capillary-dominated ($G=0$) and $a_0\sim t^{0.762}$ for gravity-dominated (large $G$) limits as $a_0\rightarrow 0$ for $V=1$, $\Omega^2=0$.  \label{fig:collapseScale}}
\end{figure}

In Figure~\ref{fig:timeG}, we numerically integrate (\ref{spreadlawS}) to show that gravity promotes hole collapse. An examination of the contact-line law (\ref{CLBCs}) reveals the mechanism behind the enhanced spreading rate; an increase in contact-angle leads to increased contact-line speed (equivalently, spreading rate).  The collapse dynamics occurs in two phases characterized by i) an initial slow spreading process that ii) accelerates as the hole collapses $a\rightarrow 0$. We plot the collapse time $t_c$, computed from (\ref{collapsetime}), against Bond number $G$ and initial contact-line radius $a_0$ in Figure~\ref{fig:collapseSessile}. For large holes, the collapse time depends strongly upon $G$. In contrast, for small holes, the collapse time appears to be independent of $G$ consistent with the relative increase in importance of surface tension forces at small scales. In Figure~\ref{fig:collapseScale}, we plot initial contact-line radius $a_0$ against collapse time $t_c$ to show the large $G$ limit exhibits power law behavior $a\sim t^{0.762}$, whose exponent is identical to that reported by \textcite{diez1992self} for converging viscous gravity currents. This is a limiting case (large $G$) of our model. When combined with our prediction for the asymptotic behavior for $G=0$, we see that our model has wide applicability from the capillary- to gravity-dominant limits.  %In the small hole limit $a_0\rightarrow 0$, the collapse dynamics follows the power law $a_0\sim t^{0.55}$, as shown in Figure~\ref{fig:collapseScale}. %This particular power law form has recently been reported in experiments on hole collapse by \textcite{dijksman2014self}. Furthermore, we can take the asymptotic limit $a_0\rightarrow 0$ of the capillary-dominant collapse time (\ref{tcG0}),
%\begin{equation} \lambda t_c \sim \frac{\pi}{8V}\left(4\ln(a_0^{-1})-1\right)a_0^2,\label{collapsesmall}\end{equation}
%to generate a lower bound on the power law and a more specific functional form to the asymptotic collapse dynamics. As could be expected, the dynamics are governed by surface tension forces in the small hole asymptotic limit.

%\begin{equation}\lambda t_c \sim \frac{\pi}{4V} \left(1-2\gamma - 4/G^2 + 2 K_1(G)/I_1(G) + \ln(4) - 2\ln\left(a_0 G\right)\right) a_0^2  \end{equation}
%with $\gamma\approx 0.557216$ Euler's constant

%\subsubsection{Volume effects}
%rescale with height $H$ predictions compare with experiment \textcite[Fig. 5b]{dijksman2014self}

Finally, we explicitly show the capillary-dominated collapse time (\ref{tcG0}) is inversely proportional to the volume $V$. Since both capillary (\ref{solG0}) and gravity-dominated (\ref{solG}) solutions are linear in $V$, the collapse time when gravitational effects are included should also be inversely proportional to $V$.  We can connect our results to experiment by choosing to scale length with the film height $\bar{h}$, instead of the lateral support radius $R$, which results in $t_c \approx \bar{h}^{-3}$ consistent with experimental observations \textcite[Fig. 5b]{dijksman2014self}. These comparisons further demonstrate the validity of our model.

\subsection{Rotational effects}

\begin{figure*}
\begin{center}
\begin{tabular}{ll}
($a$) & ($b$) \\
\includegraphics[width=0.4\textwidth]{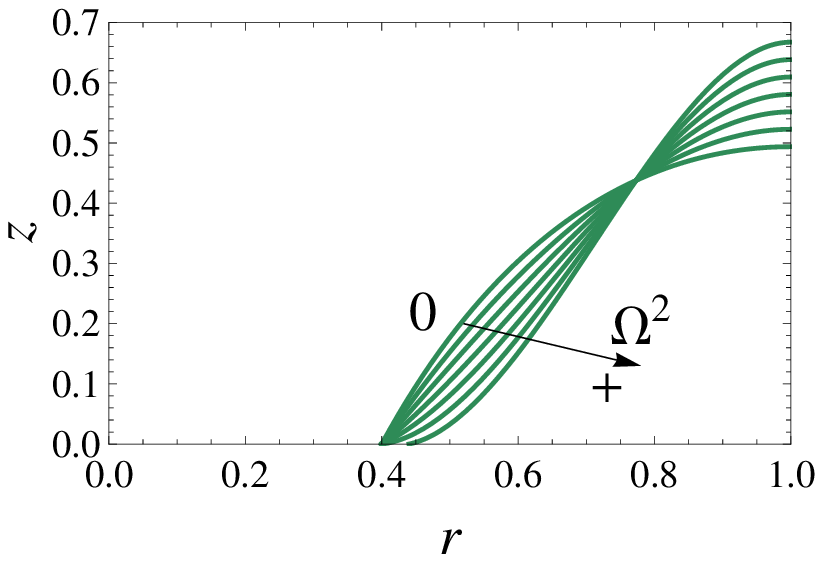}&
\includegraphics[width=0.4\textwidth]{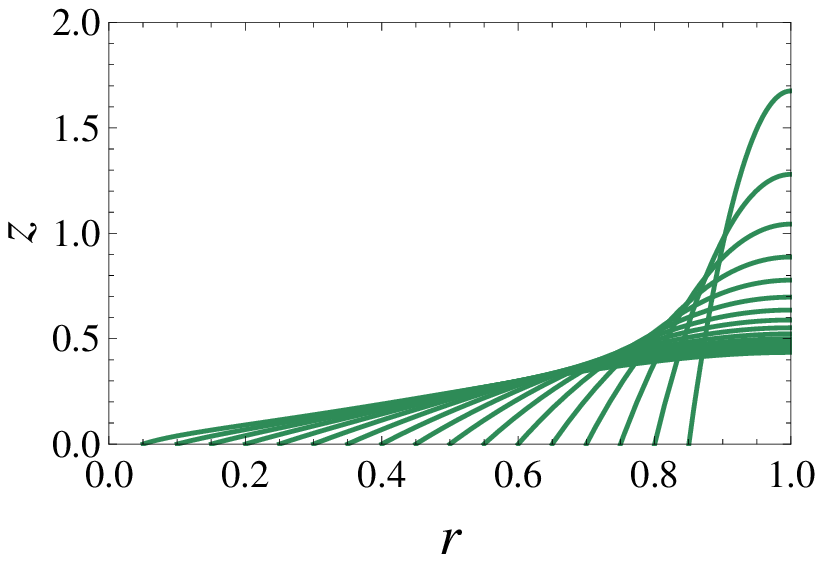}
\end{tabular}
\end{center}
\caption{Interface shapes in a rotating geometry ($G=0,V=1$): ($a$) fixed contact-line radius $a$ and varying centrifugal number $\Omega^2$ and ($b$) fixed $\Omega^2=10$ and varying $a$.  \label{fig:equilROT}}
\end{figure*}

\begin{figure}
\begin{center}
\includegraphics[width=0.4\textwidth]{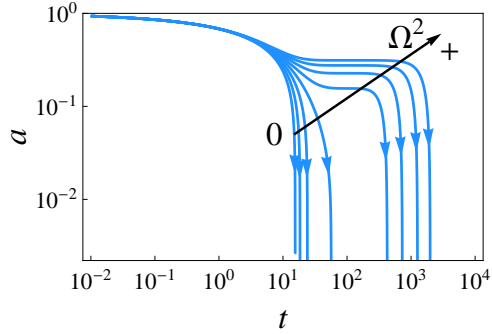}
\end{center}
\caption{Evolution of the contact-line radius $a$ against time $t$ for initial conditions $a_0=0.99$, with $V=1$, $G=0$, as it depends upon centrifugal number $\Omega^2$. \label{fig:evolutionROT}}
\end{figure}

An initially flat thin film in a rotating geometry can be made to dewet the substrate at the axis-of-rotation ($r=0$), thereby creating a hole, provided $\Omega^2\geq \Omega^2_c \equiv 48 V/\pi$ (see Appendix). This occurs, of course, because centrifugal forces tend to drive fluid to the edge of the rotating bucket. We are interested in how centrifugal forces affect the collapse dynamics of a pre-nucleated hole with radius $a_0>0$. For simplicity, we focus on a hole in a rotating geometry with $G=0$ and $\Omega^2<\Omega^2_c$. In this case, the solution of (\ref{evoleq}) is given by
\begin{widetext}
\begin{equation}\begin{split}h(r)=\left(\left(a^2-r^2\right)\left(192V+\Omega^2\pi(a^2-1)(-14+7a^2+a^4-3r^2(a^2-3))\right) \right.\\ \left. -4\ln(a)\left(96V+\Omega^2\pi\left(-2-3a^4+2a^6+6r^2-3r^4\right)\right) \right.\\ \left. +
8\ln(r)\left(48V+\Omega^2\pi(a^2-1)^3\right)\right)/(96\pi\left(3-4a^2+a^4+4\ln(a)\right)).\end{split}\label{equilROT}\end{equation}
\end{widetext}
Figure~\ref{fig:equilROT} plots typical solutions. In Figure~\ref{fig:equilROT}($a$), we show that increasing the rotation rate $\Omega^2$ tends to i) move fluid towards the edge of the bucket ($r=1$) and ii) decrease the contact angle, for fixed contact-line radius $a$. Hence, we expect centrifugal forces to slow the collapse rate with mechanism consistent with the contact-line law (\ref{CLBCs}). Figure~\ref{fig:evolutionROT} plots the evolution of the fluid interface during the collapse process for $\Omega^2=10$ showing that the contact-angle decreases as the contact-line radius $a$ decreases.

In situations where the fluid hole is rotating with $\Omega^2<\Omega^2_c$, centrifugal forces can slow down the collapse process by decreasing the contact-angle and therefore the contact-line speed according to (\ref{CLBCs}). Figure~\ref{fig:evolutionROT} plots the evolution of the contact-line radius $a$ against time, as it depends upon $\Omega^2$. For increasing $\Omega^2$, the spreading dynamics become more complex as witnessed by the pronounced plateau, characterized by stalled spreading behavior, that separates the large $a$ and small $a$ regions. In the plateau region, the contact-angle approaches zero leading to slow collapse dynamics for a finite period of time until surface tension forces become dominant and control the dynamics according to the asymptotics previously discussed, Eq. (\ref{collapsesmall}). Note the size of the plateau and the range of the small $a$ region both increase with $\Omega^2$. This implies that the rotation rate could be used as an effective mechanism to control the collapse dynamics in practice.

%\begin{figure}
%\begin{tabular}{l}
%($a$)\\
%\includegraphics[width=0.4\textwidth]{equilibriaRotate2}\\
% ($b$) \\
%\includegraphics[width=0.4\textwidth]{equilibriaR10}%
%\end{tabular}
%\caption{Interface shapes in a rotating geometry ($G=0,V=1$): ($a$) fixed contact-line radius $a$ and varying centrifugal number $\Omega^2$ and ($b$) fixed $\Omega^2=10$ and varying $a$.  \label{fig:equilROT}}
%\end{figure}

\begin{figure}
\begin{center}
\includegraphics[width=0.4\textwidth]{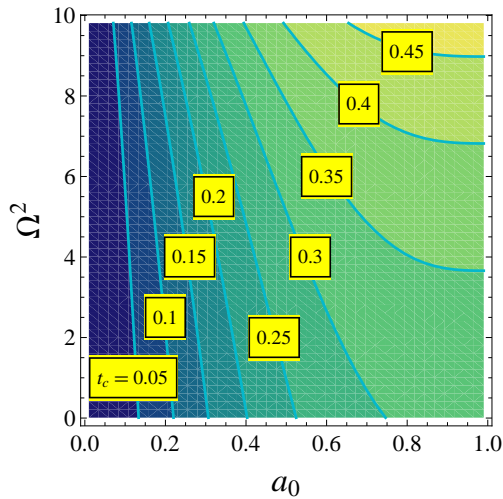}
\end{center}
\caption{ Collapse time $t_c$ against centrifugal number $\Omega^2$ and initial contact-line radius $a_0$ for $V=1$, $G=0$.  \label{fig:collapseROT}}
\end{figure}

As we have shown, centrifugal forces can dramatically slow down the spreading speed of a fluid hole through the mechanics of the contact-line speed law (\ref{spreadlawS}). Figure~\ref{fig:collapseROT} plots the collapse time $t_c$ against centrifugal number $\Omega^2$ and initial contact-line radius $a_0$, showing that centrifugal forces are more effective at increasing the total collapse time for large initial holes. In contrast, the total collapse time is insensitive to centrifugal forces $\Omega^2$ for small initial holes, because surface tension forces dominate the collapse dynamics in this limit. This observation was also true for gravitational forces and appears to be universal.

\subsection{Experimental comparison: volume effects}
\begin{figure*}
\begin{center}
\begin{tabular}{ll}
($a$) & ($b$) \\
\includegraphics[width=0.42\textwidth]{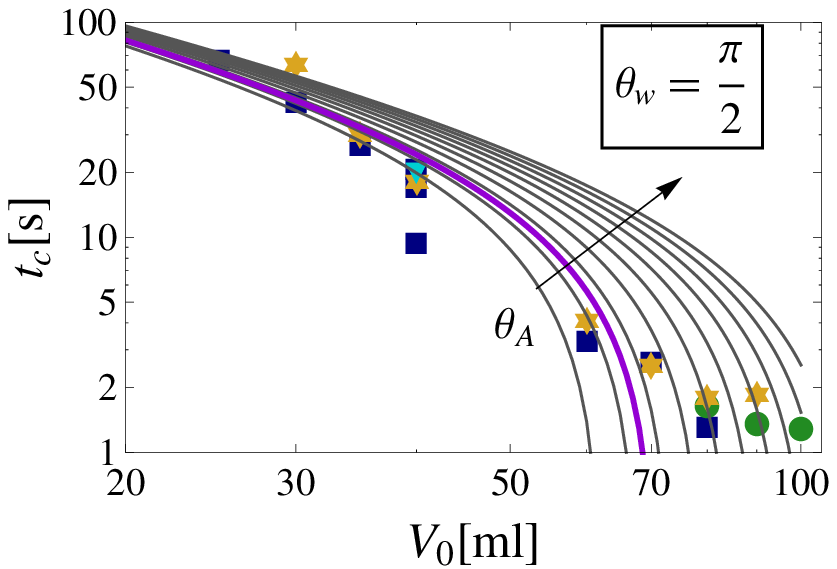}&
\includegraphics[width=0.42\textwidth]{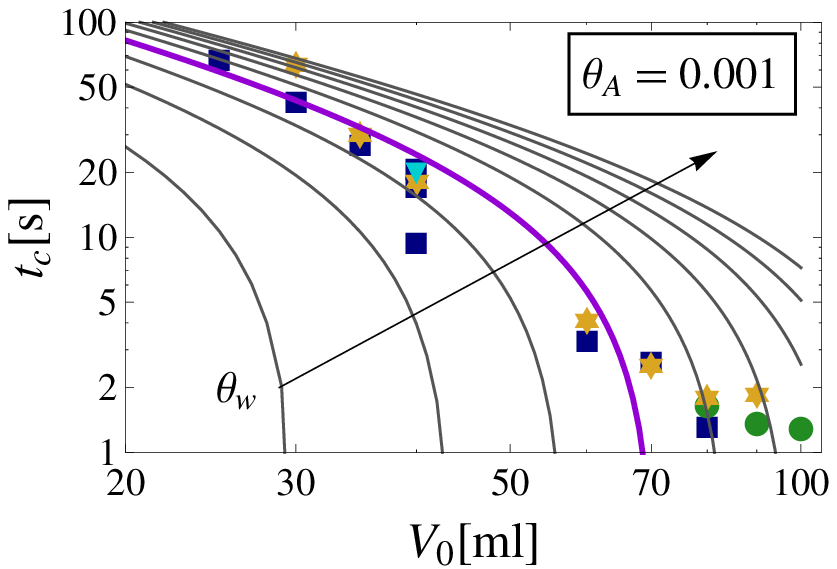}
\end{tabular}
\end{center}
\caption{Collapse time $t_c$[s] against volume $V_0$[ml] for ($a$) fixed $\theta_w=\pi/2$, varying $\theta_A$ and ($b$) fixed $\theta_A=0.001$, varying $\theta_w$. Symbols correspond to experimental data with pre-spin rotation rates $\omega=1-1.7$~rps (blue $\square$: 1.0~rps, yellow $\star$ 1.2~rps, green $\bullet$ 1.5~rps, cyan $\triangledown$ 1.7~rps) and $\mu=1000$~mPa$\cdot$s, $\sigma=0.02$~N/m, $R=6.5$~cm. The best fit to the 1.0~rps (blue $\square$) data set is shown with thick line-type and corresponds to $\Lambda=0.105$mm/s, $\theta_w=\pi/2$ and $\theta_A=0.001$. \label{fig:volumeEXP}}
\end{figure*}
We can test our model further by comparing with experiments whose protocol is described in detail in \textcite{dijksman2014self}. In our experiments an initial thin film is spun into a fluid hole; we fix the volume $V$ and create a different $a_0(V,\Omega)$ by choosing different pre-collapse $\Omega$. Note the collapse proceeds once rotation stops so that $\Omega=0$ throughout the experiment. This allows us to also check the predictions for $t_c(a_0)$. Our experimental collapse preparations creates a nonlinear relation for $a_0(V,\Omega)$, which also depends on $\theta_A$ and $\theta_w$ (wall contact-angle). In fact, the latter dependency is quite strong, as the volume integral for $h(r)$ is weighted proportionally to the radius. Nevertheless, we can extract $t_c(a_0(V,\Omega))$ for a range of volumes and preparatory rotation rates $\Omega$. We perform these experiments by imaging the collapse process from above with a simple digital camera \citep[Figs. 2a,b]{dijksman2014self}. The volumes explored are 25 to 100ml; the range for $\omega$ is 1-1.7~rps. We create initial conditions by pre-spinning the container for 60-100 seconds for all experiments. Unfortunately the applied experimental procedure does not allow us to directly measure $a_0(V,\Omega)$, but we can plot the observed $t_c$ against a family of curves computed for a reasonable range of $\theta_A$ and $\theta_w$ (Figure~\ref{fig:volumeEXP}). The fit parameter $\Lambda$ allows for a vertical scaling of the curves. Despite the relative freedom in choosing $\theta_A$, $\theta_w$ and $\Lambda$, we conclude that also the independent $t_c$ data are consistent with our model. The best fit for the $\omega=1.0$~rps data set is shown in thick line-type in Figure~\ref{fig:volumeEXP} and corresponds to $\Lambda=0.105$mm/s, $\theta_w=\pi/2$ and $\theta_A=0.001$. At large $V_0$, there are some significant deviations; we attribute these to `waiting time' effects that are not captured by our model \citep{lacey1982waiting,marino1996waiting}.

\section{Concluding remarks}
We have studied the collapse dynamics of a nucleated hole in a thin film on a completely-wetting substrate. Our focus is the mesocopic regime, as we develop a model that encompasses both surface tension driven flows and viscous gravity currents. The model predicts a power law exponent for the time-dependent hole radius that agrees with experiment in both the capillary \citep{dijksman2014self} and gravity \citep{diez1992self} limits. Furthermore, our predictions compare favorably to these experiments over a range of volumes. We also show that centrifugal forces from the rotating geometry can lead to complex spreading dynamics, characterized by stalled spreading behavior that separates the large and small hole limits. We believe our model can help bridge the gap between the well-studied nanoscopic and macroscopic regimes to the less well-understood mesoscopic regime relevant to industrial coating processes, such as immersion lithography. For example, in coating processes fluid holes can be viewed as defects which naturally disappear on the time scale predicted by our analysis.

Our model is concerned with holes on completely-wetting substrates that must be nucleated by external means. That is, each hole we consider will always collapse. However, it is possible to have a finite-sized equilibrium hole on a partially-wetting substrate. This reflects the competition between capillarity, which drives collapse, and surface chemistry (wetting effects) that resists this motion. Can other driving forces lead to finite-size holes on a completely wetting substrate? Possible forces could include centrifugal forces, which tend to slow hole collapse, or Marangoni (thermocapillary) forces from applied thermal fields \citep{mukho11,bostwick2013thermal}. Lastly, a spreading contact-line is susceptible to fingering instabilities that are outside the scope of this paper. We plan to extend our results into these directions.

\section*{Acknowledgements}
This work was supported by the National Science Foundation under grant numbers DMS-0968258, DMS-1517291. The authors wish to thank Karen Daniels for helpful discussions surrounding this work.

\section*{Appendix}
For a film without a contact-line, we replace (\ref{BCwall},\ref{contact}) with the following boundary conditions,
\begin{equation}h'(0)=h'(1)=v(1)=0,  \end{equation}
and solve the equilibrium equation (\ref{quasistaticeq}) to yield
\begin{equation} h(r)=\frac{V}{\pi} - \frac{\Omega^2}{96}\left(2-6r^2+3r^4\right). \label{filmsol}\end{equation}
The film dewets the substrate $h=0$ along the axis-of-rotation $r=0$ at a critical centrifugal number $\Omega_c^2=48V/\pi$.

\bibliographystyle{apsrev4-1}
\bibliography{jfmHole}% Produces the bibliography via BibTeX.

\begin{thebibliography}{53}%
\makeatletter
\providecommand \@ifxundefined [1]{%
 \@ifx{#1\undefined}
}%
\providecommand \@ifnum [1]{%
 \ifnum #1\expandafter \@firstoftwo
 \else \expandafter \@secondoftwo
 \fi
}%
\providecommand \@ifx [1]{%
 \ifx #1\expandafter \@firstoftwo
 \else \expandafter \@secondoftwo
 \fi
}%
\providecommand \natexlab [1]{#1}%
\providecommand \enquote  [1]{``#1''}%
\providecommand \bibnamefont  [1]{#1}%
\providecommand \bibfnamefont [1]{#1}%
\providecommand \citenamefont [1]{#1}%
\providecommand \href@noop [0]{\@secondoftwo}%
\providecommand \href [0]{\begingroup \@sanitize@url \@href}%
\providecommand \@href[1]{\@@startlink{#1}\@@href}%
\providecommand \@@href[1]{\endgroup#1\@@endlink}%
\providecommand \@sanitize@url [0]{\catcode `\\12\catcode `\$12\catcode
  `\&12\catcode `\#12\catcode `\^12\catcode `\_12\catcode `\%12\relax}%
\providecommand \@@startlink[1]{}%
\providecommand \@@endlink[0]{}%
\providecommand \url  [0]{\begingroup\@sanitize@url \@url }%
\providecommand \@url [1]{\endgroup\@href {#1}{\urlprefix }}%
\providecommand \urlprefix  [0]{URL }%
\providecommand \Eprint [0]{\href }%
\providecommand \doibase [0]{http://dx.doi.org/}%
\providecommand \selectlanguage [0]{\@gobble}%
\providecommand \bibinfo  [0]{\@secondoftwo}%
\providecommand \bibfield  [0]{\@secondoftwo}%
\providecommand \translation [1]{[#1]}%
\providecommand \BibitemOpen [0]{}%
\providecommand \bibitemStop [0]{}%
\providecommand \bibitemNoStop [0]{.\EOS\space}%
\providecommand \EOS [0]{\spacefactor3000\relax}%
\providecommand \BibitemShut  [1]{\csname bibitem#1\endcsname}%
\let\auto@bib@innerbib\@empty
%</preamble>
\bibitem [{\citenamefont {Martin}\ \emph {et~al.}(2000)\citenamefont {Martin},
  \citenamefont {Rossier}, \citenamefont {Buguin}, \citenamefont {Auroy},\ and\
  \citenamefont {Brochard-Wyart}}]{martin2000spinodal}%
  \BibitemOpen
  \bibfield  {author} {\bibinfo {author} {\bibfnamefont {A.}~\bibnamefont
  {Martin}}, \bibinfo {author} {\bibfnamefont {O.}~\bibnamefont {Rossier}},
  \bibinfo {author} {\bibfnamefont {A.}~\bibnamefont {Buguin}}, \bibinfo
  {author} {\bibfnamefont {P.}~\bibnamefont {Auroy}}, \ and\ \bibinfo {author}
  {\bibfnamefont {F.}~\bibnamefont {Brochard-Wyart}},\ }\href@noop {}
  {\bibfield  {journal} {\bibinfo  {journal} {The European Physical Journal E}\
  }\textbf {\bibinfo {volume} {3}},\ \bibinfo {pages} {337} (\bibinfo {year}
  {2000})}\BibitemShut {NoStop}%
\bibitem [{\citenamefont {Gentili}\ \emph {et~al.}(2012)\citenamefont
  {Gentili}, \citenamefont {Foschi}, \citenamefont {Valle}, \citenamefont
  {Cavallini},\ and\ \citenamefont {Biscarini}}]{gentili2012applications}%
  \BibitemOpen
  \bibfield  {author} {\bibinfo {author} {\bibfnamefont {D.}~\bibnamefont
  {Gentili}}, \bibinfo {author} {\bibfnamefont {G.}~\bibnamefont {Foschi}},
  \bibinfo {author} {\bibfnamefont {F.}~\bibnamefont {Valle}}, \bibinfo
  {author} {\bibfnamefont {M.}~\bibnamefont {Cavallini}}, \ and\ \bibinfo
  {author} {\bibfnamefont {F.}~\bibnamefont {Biscarini}},\ }\href@noop {}
  {\bibfield  {journal} {\bibinfo  {journal} {Chemical Society Reviews}\
  }\textbf {\bibinfo {volume} {41}},\ \bibinfo {pages} {4430} (\bibinfo {year}
  {2012})}\BibitemShut {NoStop}%
\bibitem [{\citenamefont {Ferrer}\ \emph {et~al.}(2014)\citenamefont {Ferrer},
  \citenamefont {Halajko},\ and\ \citenamefont {Amatucci}}]{ferrer2014micro}%
  \BibitemOpen
  \bibfield  {author} {\bibinfo {author} {\bibfnamefont {A.~J.}\ \bibnamefont
  {Ferrer}}, \bibinfo {author} {\bibfnamefont {A.}~\bibnamefont {Halajko}}, \
  and\ \bibinfo {author} {\bibfnamefont {G.~G.}\ \bibnamefont {Amatucci}},\
  }\href@noop {} {\bibfield  {journal} {\bibinfo  {journal} {Advanced
  Engineering Materials}\ }\textbf {\bibinfo {volume} {16}},\ \bibinfo {pages}
  {1167} (\bibinfo {year} {2014})}\BibitemShut {NoStop}%
\bibitem [{\citenamefont {Martin}\ \emph {et~al.}(2001)\citenamefont {Martin},
  \citenamefont {Buguin},\ and\ \citenamefont
  {Brochard-Wyart}}]{martin2001dewetting}%
  \BibitemOpen
  \bibfield  {author} {\bibinfo {author} {\bibfnamefont {A.}~\bibnamefont
  {Martin}}, \bibinfo {author} {\bibfnamefont {A.}~\bibnamefont {Buguin}}, \
  and\ \bibinfo {author} {\bibfnamefont {F.}~\bibnamefont {Brochard-Wyart}},\
  }\href@noop {} {\bibfield  {journal} {\bibinfo  {journal} {Langmuir}\
  }\textbf {\bibinfo {volume} {17}},\ \bibinfo {pages} {6553} (\bibinfo {year}
  {2001})}\BibitemShut {NoStop}%
\bibitem [{\citenamefont {Geoghegan}\ and\ \citenamefont
  {Krausch}(2003)}]{geoghegan2003wetting}%
  \BibitemOpen
  \bibfield  {author} {\bibinfo {author} {\bibfnamefont {M.}~\bibnamefont
  {Geoghegan}}\ and\ \bibinfo {author} {\bibfnamefont {G.}~\bibnamefont
  {Krausch}},\ }\href@noop {} {\bibfield  {journal} {\bibinfo  {journal}
  {Progress in Polymer Science}\ }\textbf {\bibinfo {volume} {28}},\ \bibinfo
  {pages} {261} (\bibinfo {year} {2003})}\BibitemShut {NoStop}%
\bibitem [{\citenamefont {Sellier}\ \emph {et~al.}(2015)\citenamefont
  {Sellier}, \citenamefont {Grayson}, \citenamefont {Renbaum-Wolff},
  \citenamefont {Song},\ and\ \citenamefont {Bertram}}]{sellier2015estimating}%
  \BibitemOpen
  \bibfield  {author} {\bibinfo {author} {\bibfnamefont {M.}~\bibnamefont
  {Sellier}}, \bibinfo {author} {\bibfnamefont {J.}~\bibnamefont {Grayson}},
  \bibinfo {author} {\bibfnamefont {L.}~\bibnamefont {Renbaum-Wolff}}, \bibinfo
  {author} {\bibfnamefont {M.}~\bibnamefont {Song}}, \ and\ \bibinfo {author}
  {\bibfnamefont {A.}~\bibnamefont {Bertram}},\ }\href@noop {} {\bibfield
  {journal} {\bibinfo  {journal} {Journal of Rheology (1978-present)}\ }\textbf
  {\bibinfo {volume} {59}},\ \bibinfo {pages} {733} (\bibinfo {year}
  {2015})}\BibitemShut {NoStop}%
\bibitem [{\citenamefont {Redon}\ \emph {et~al.}(1991)\citenamefont {Redon},
  \citenamefont {Brochard-Wyart},\ and\ \citenamefont {Rondelez}}]{redon1991}%
  \BibitemOpen
  \bibfield  {author} {\bibinfo {author} {\bibfnamefont {C.}~\bibnamefont
  {Redon}}, \bibinfo {author} {\bibfnamefont {F.}~\bibnamefont
  {Brochard-Wyart}}, \ and\ \bibinfo {author} {\bibfnamefont {F.}~\bibnamefont
  {Rondelez}},\ }\href {\doibase 10.1103/PhysRevLett.66.715} {\bibfield
  {journal} {\bibinfo  {journal} {Phys. Rev. Lett.}\ }\textbf {\bibinfo
  {volume} {66}},\ \bibinfo {pages} {715} (\bibinfo {year} {1991})}\BibitemShut
  {NoStop}%
\bibitem [{\citenamefont {Padday}(1970)}]{padday1970cohesive}%
  \BibitemOpen
  \bibfield  {author} {\bibinfo {author} {\bibfnamefont {J.}~\bibnamefont
  {Padday}},\ }\href@noop {} {\bibfield  {journal} {\bibinfo  {journal}
  {Special Discussions of the Faraday Society}\ }\textbf {\bibinfo {volume}
  {1}},\ \bibinfo {pages} {64} (\bibinfo {year} {1970})}\BibitemShut {NoStop}%
\bibitem [{\citenamefont {Taylor}\ and\ \citenamefont
  {Michael}(1973)}]{taylor1973making}%
  \BibitemOpen
  \bibfield  {author} {\bibinfo {author} {\bibfnamefont {G.}~\bibnamefont
  {Taylor}}\ and\ \bibinfo {author} {\bibfnamefont {D.}~\bibnamefont
  {Michael}},\ }\href@noop {} {\bibfield  {journal} {\bibinfo  {journal}
  {Journal of fluid mechanics}\ }\textbf {\bibinfo {volume} {58}},\ \bibinfo
  {pages} {625} (\bibinfo {year} {1973})}\BibitemShut {NoStop}%
\bibitem [{\citenamefont {Dhiman}\ and\ \citenamefont
  {Chandra}(2009)}]{dhiman2009rupture}%
  \BibitemOpen
  \bibfield  {author} {\bibinfo {author} {\bibfnamefont {R.}~\bibnamefont
  {Dhiman}}\ and\ \bibinfo {author} {\bibfnamefont {S.}~\bibnamefont
  {Chandra}},\ }in\ \href@noop {} {\emph {\bibinfo {booktitle} {Proceedings of
  the Royal Society of London A: Mathematical, Physical and Engineering
  Sciences}}}\ (\bibinfo {organization} {The Royal Society},\ \bibinfo {year}
  {2009})\ p.\ \bibinfo {pages} {rspa20090425}\BibitemShut {NoStop}%
\bibitem [{\citenamefont {Diez}\ \emph {et~al.}(1992)\citenamefont {Diez},
  \citenamefont {Gratton},\ and\ \citenamefont {Gratton}}]{diez1992self}%
  \BibitemOpen
  \bibfield  {author} {\bibinfo {author} {\bibfnamefont {J.~A.}\ \bibnamefont
  {Diez}}, \bibinfo {author} {\bibfnamefont {R.}~\bibnamefont {Gratton}}, \
  and\ \bibinfo {author} {\bibfnamefont {J.}~\bibnamefont {Gratton}},\
  }\href@noop {} {\bibfield  {journal} {\bibinfo  {journal} {Physics of Fluids
  A: Fluid Dynamics (1989-1993)}\ }\textbf {\bibinfo {volume} {4}},\ \bibinfo
  {pages} {1148} (\bibinfo {year} {1992})}\BibitemShut {NoStop}%
\bibitem [{\citenamefont {Backholm}\ \emph {et~al.}(2014)\citenamefont
  {Backholm}, \citenamefont {Benzaquen}, \citenamefont {Salez}, \citenamefont
  {Rapha{\"e}l},\ and\ \citenamefont {Dalnoki-Veress}}]{backholm2014capillary}%
  \BibitemOpen
  \bibfield  {author} {\bibinfo {author} {\bibfnamefont {M.}~\bibnamefont
  {Backholm}}, \bibinfo {author} {\bibfnamefont {M.}~\bibnamefont {Benzaquen}},
  \bibinfo {author} {\bibfnamefont {T.}~\bibnamefont {Salez}}, \bibinfo
  {author} {\bibfnamefont {E.}~\bibnamefont {Rapha{\"e}l}}, \ and\ \bibinfo
  {author} {\bibfnamefont {K.}~\bibnamefont {Dalnoki-Veress}},\ }\href@noop {}
  {\bibfield  {journal} {\bibinfo  {journal} {Soft Matter}\ }\textbf {\bibinfo
  {volume} {10}},\ \bibinfo {pages} {2550} (\bibinfo {year}
  {2014})}\BibitemShut {NoStop}%
\bibitem [{\citenamefont {Mukhopadhyay}\ and\ \citenamefont
  {Behringer}(2009)}]{mukho09}%
  \BibitemOpen
  \bibfield  {author} {\bibinfo {author} {\bibfnamefont {S.}~\bibnamefont
  {Mukhopadhyay}}\ and\ \bibinfo {author} {\bibfnamefont {R.}~\bibnamefont
  {Behringer}},\ }\href@noop {} {\bibfield  {journal} {\bibinfo  {journal} {J.
  Physics: Condensed Matter}\ }\textbf {\bibinfo {volume} {21}},\ \bibinfo
  {pages} {464123} (\bibinfo {year} {2009})}\BibitemShut {NoStop}%
\bibitem [{\citenamefont {Dijksman}\ \emph {et~al.}(2015)\citenamefont
  {Dijksman}, \citenamefont {Mukhopadhyay}, \citenamefont {Gaebler},
  \citenamefont {Witelski},\ and\ \citenamefont
  {Behringer}}]{dijksman2014self}%
  \BibitemOpen
  \bibfield  {author} {\bibinfo {author} {\bibfnamefont {J.~A.}\ \bibnamefont
  {Dijksman}}, \bibinfo {author} {\bibfnamefont {S.}~\bibnamefont
  {Mukhopadhyay}}, \bibinfo {author} {\bibfnamefont {C.}~\bibnamefont
  {Gaebler}}, \bibinfo {author} {\bibfnamefont {T.~P.}\ \bibnamefont
  {Witelski}}, \ and\ \bibinfo {author} {\bibfnamefont {R.~P.}\ \bibnamefont
  {Behringer}},\ }\href@noop {} {\bibfield  {journal} {\bibinfo  {journal}
  {Physical Review E}\ }\textbf {\bibinfo {volume} {92}},\ \bibinfo {pages}
  {043016} (\bibinfo {year} {2015})}\BibitemShut {NoStop}%
\bibitem [{\citenamefont {Sharma}\ and\ \citenamefont
  {Ruckenstein}(1990)}]{sharma1990energetic}%
  \BibitemOpen
  \bibfield  {author} {\bibinfo {author} {\bibfnamefont {A.}~\bibnamefont
  {Sharma}}\ and\ \bibinfo {author} {\bibfnamefont {E.}~\bibnamefont
  {Ruckenstein}},\ }\href@noop {} {\bibfield  {journal} {\bibinfo  {journal}
  {Journal of Colloid and Interface Science}\ }\textbf {\bibinfo {volume}
  {137}},\ \bibinfo {pages} {433} (\bibinfo {year} {1990})}\BibitemShut
  {NoStop}%
\bibitem [{\citenamefont {Moriarty}\ and\ \citenamefont
  {Schwartz}(1993)}]{moriarty1993dynamic}%
  \BibitemOpen
  \bibfield  {author} {\bibinfo {author} {\bibfnamefont {J.}~\bibnamefont
  {Moriarty}}\ and\ \bibinfo {author} {\bibfnamefont {L.}~\bibnamefont
  {Schwartz}},\ }\href@noop {} {\bibfield  {journal} {\bibinfo  {journal}
  {Journal of colloid and interface science}\ }\textbf {\bibinfo {volume}
  {161}},\ \bibinfo {pages} {335} (\bibinfo {year} {1993})}\BibitemShut
  {NoStop}%
\bibitem [{\citenamefont {Bankoff}\ \emph {et~al.}(2003)\citenamefont
  {Bankoff}, \citenamefont {Johnson}, \citenamefont {Miksis}, \citenamefont
  {Schulter},\ and\ \citenamefont {Lopez}}]{bankoffdryspot}%
  \BibitemOpen
  \bibfield  {author} {\bibinfo {author} {\bibfnamefont {S.~G.}\ \bibnamefont
  {Bankoff}}, \bibinfo {author} {\bibfnamefont {M.~F.~G.}\ \bibnamefont
  {Johnson}}, \bibinfo {author} {\bibfnamefont {M.~J.}\ \bibnamefont {Miksis}},
  \bibinfo {author} {\bibfnamefont {R.~A.}\ \bibnamefont {Schulter}}, \ and\
  \bibinfo {author} {\bibfnamefont {P.~G.}\ \bibnamefont {Lopez}},\ }\href
  {\doibase 10.1017/S0022112003004634} {\bibfield  {journal} {\bibinfo
  {journal} {Journal of Fluid Mechanics}\ }\textbf {\bibinfo {volume} {486}},\
  \bibinfo {pages} {239} (\bibinfo {year} {2003})}\BibitemShut {NoStop}%
\bibitem [{\citenamefont {L{\'o}pez}\ \emph {et~al.}(2001)\citenamefont
  {L{\'o}pez}, \citenamefont {Miksis},\ and\ \citenamefont
  {Bankoff}}]{lopez2001stability}%
  \BibitemOpen
  \bibfield  {author} {\bibinfo {author} {\bibfnamefont {P.}~\bibnamefont
  {L{\'o}pez}}, \bibinfo {author} {\bibfnamefont {M.}~\bibnamefont {Miksis}}, \
  and\ \bibinfo {author} {\bibfnamefont {S.}~\bibnamefont {Bankoff}},\
  }\href@noop {} {\bibfield  {journal} {\bibinfo  {journal} {Physics of Fluids
  (1994-present)}\ }\textbf {\bibinfo {volume} {13}},\ \bibinfo {pages} {1601}
  (\bibinfo {year} {2001})}\BibitemShut {NoStop}%
\bibitem [{\citenamefont {Lubarda}(2013)}]{lubarda2013shape}%
  \BibitemOpen
  \bibfield  {author} {\bibinfo {author} {\bibfnamefont {V.~A.}\ \bibnamefont
  {Lubarda}},\ }\href@noop {} {\bibfield  {journal} {\bibinfo  {journal} {Acta
  Mechanica}\ }\textbf {\bibinfo {volume} {224}},\ \bibinfo {pages} {1365}
  (\bibinfo {year} {2013})}\BibitemShut {NoStop}%
\bibitem [{\citenamefont {Ungarish}(2009)}]{ungarish2009book}%
  \BibitemOpen
  \bibfield  {author} {\bibinfo {author} {\bibfnamefont {M.}~\bibnamefont
  {Ungarish}},\ }\href {https://books.google.com/books?id=tM\_44bTSCUQC} {\emph
  {\bibinfo {title} {An Introduction to Gravity Currents and Intrusions}}}\
  (\bibinfo  {publisher} {CRC Press},\ \bibinfo {year} {2009})\BibitemShut
  {NoStop}%
\bibitem [{\citenamefont {Huppert}(1986)}]{huppert1986intrusion}%
  \BibitemOpen
  \bibfield  {author} {\bibinfo {author} {\bibfnamefont {H.~E.}\ \bibnamefont
  {Huppert}},\ }\href@noop {} {\bibfield  {journal} {\bibinfo  {journal}
  {Journal of Fluid Mechanics}\ }\textbf {\bibinfo {volume} {173}},\ \bibinfo
  {pages} {557} (\bibinfo {year} {1986})}\BibitemShut {NoStop}%
\bibitem [{\citenamefont {Huppert}(1982)}]{huppert1982propagation}%
  \BibitemOpen
  \bibfield  {author} {\bibinfo {author} {\bibfnamefont {H.~E.}\ \bibnamefont
  {Huppert}},\ }\href@noop {} {\bibfield  {journal} {\bibinfo  {journal}
  {Journal of Fluid Mechanics}\ }\textbf {\bibinfo {volume} {121}},\ \bibinfo
  {pages} {43} (\bibinfo {year} {1982})}\BibitemShut {NoStop}%
\bibitem [{\citenamefont {Gratton}\ and\ \citenamefont
  {Minotti}(1990)}]{gratton1990self}%
  \BibitemOpen
  \bibfield  {author} {\bibinfo {author} {\bibfnamefont {J.}~\bibnamefont
  {Gratton}}\ and\ \bibinfo {author} {\bibfnamefont {F.}~\bibnamefont
  {Minotti}},\ }\href@noop {} {\bibfield  {journal} {\bibinfo  {journal}
  {Journal of Fluid Mechanics}\ }\textbf {\bibinfo {volume} {210}},\ \bibinfo
  {pages} {155} (\bibinfo {year} {1990})}\BibitemShut {NoStop}%
\bibitem [{\citenamefont {Ancey}\ \emph {et~al.}(2009)\citenamefont {Ancey},
  \citenamefont {Cochard},\ and\ \citenamefont {Andreini}}]{ancey2009dam}%
  \BibitemOpen
  \bibfield  {author} {\bibinfo {author} {\bibfnamefont {C.}~\bibnamefont
  {Ancey}}, \bibinfo {author} {\bibfnamefont {S.}~\bibnamefont {Cochard}}, \
  and\ \bibinfo {author} {\bibfnamefont {N.}~\bibnamefont {Andreini}},\
  }\href@noop {} {\bibfield  {journal} {\bibinfo  {journal} {Journal of Fluid
  Mechanics}\ }\textbf {\bibinfo {volume} {624}},\ \bibinfo {pages} {1}
  (\bibinfo {year} {2009})}\BibitemShut {NoStop}%
\bibitem [{\citenamefont {Dussan~V.}(1979)}]{dussan79}%
  \BibitemOpen
  \bibfield  {author} {\bibinfo {author} {\bibfnamefont {E.}~\bibnamefont
  {Dussan~V.}},\ }\href@noop {} {\bibfield  {journal} {\bibinfo  {journal}
  {Annual Review of Fluid Mechanics}\ }\textbf {\bibinfo {volume} {11}},\
  \bibinfo {pages} {371} (\bibinfo {year} {1979})}\BibitemShut {NoStop}%
\bibitem [{\citenamefont {de~Gennes}(1985)}]{degennes85}%
  \BibitemOpen
  \bibfield  {author} {\bibinfo {author} {\bibfnamefont {P.}~\bibnamefont
  {de~Gennes}},\ }\href@noop {} {\bibfield  {journal} {\bibinfo  {journal}
  {Reviews of Modern Physics}\ }\textbf {\bibinfo {volume} {57}},\ \bibinfo
  {pages} {827} (\bibinfo {year} {1985})}\BibitemShut {NoStop}%
\bibitem [{\citenamefont {Bonn}\ \emph {et~al.}(2009)\citenamefont {Bonn},
  \citenamefont {Eggers}, \citenamefont {Indekeu}, \citenamefont {Meunier},\
  and\ \citenamefont {Rolley}}]{bonn09}%
  \BibitemOpen
  \bibfield  {author} {\bibinfo {author} {\bibfnamefont {D.}~\bibnamefont
  {Bonn}}, \bibinfo {author} {\bibfnamefont {J.}~\bibnamefont {Eggers}},
  \bibinfo {author} {\bibfnamefont {J.}~\bibnamefont {Indekeu}}, \bibinfo
  {author} {\bibfnamefont {J.}~\bibnamefont {Meunier}}, \ and\ \bibinfo
  {author} {\bibfnamefont {E.}~\bibnamefont {Rolley}},\ }\href@noop {}
  {\bibfield  {journal} {\bibinfo  {journal} {Reviews of Modern Physics}\
  }\textbf {\bibinfo {volume} {81}},\ \bibinfo {pages} {739} (\bibinfo {year}
  {2009})}\BibitemShut {NoStop}%
\bibitem [{\citenamefont {Snoeijer}\ and\ \citenamefont
  {Andreotti}(2013)}]{snoeijer}%
  \BibitemOpen
  \bibfield  {author} {\bibinfo {author} {\bibfnamefont {J.~H.}\ \bibnamefont
  {Snoeijer}}\ and\ \bibinfo {author} {\bibfnamefont {B.}~\bibnamefont
  {Andreotti}},\ }\href {\doibase 10.1146/annurev-fluid-011212-140734}
  {\bibfield  {journal} {\bibinfo  {journal} {Annual Review of Fluid
  Mechanics}\ }\textbf {\bibinfo {volume} {45}},\ \bibinfo {pages} {269}
  (\bibinfo {year} {2013})},\ \Eprint
  {http://arxiv.org/abs/http://www.annualreviews.org/doi/pdf/10.1146/annurev-f%
luid-011212-140734}
  {http://www.annualreviews.org/doi/pdf/10.1146/annurev-fluid-011212-140734}
  \BibitemShut {NoStop}%
\bibitem [{\citenamefont {Huh}\ and\ \citenamefont {Scriven}(1971)}]{huh71}%
  \BibitemOpen
  \bibfield  {author} {\bibinfo {author} {\bibfnamefont {C.}~\bibnamefont
  {Huh}}\ and\ \bibinfo {author} {\bibfnamefont {L.}~\bibnamefont {Scriven}},\
  }\href@noop {} {\bibfield  {journal} {\bibinfo  {journal} {Journal of Colloid
  and Interface Science}\ }\textbf {\bibinfo {volume} {35}},\ \bibinfo {pages}
  {85} (\bibinfo {year} {1971})}\BibitemShut {NoStop}%
\bibitem [{\citenamefont {Greenspan}(1978)}]{greenspan}%
  \BibitemOpen
  \bibfield  {author} {\bibinfo {author} {\bibfnamefont {H.}~\bibnamefont
  {Greenspan}},\ }\href@noop {} {\bibfield  {journal} {\bibinfo  {journal} {J.
  Fluid Mech.}\ }\textbf {\bibinfo {volume} {84}},\ \bibinfo {pages} {125}
  (\bibinfo {year} {1978})}\BibitemShut {NoStop}%
\bibitem [{\citenamefont {Bostwick}\ and\ \citenamefont
  {Steen}(2014)}]{bostwickS}%
  \BibitemOpen
  \bibfield  {author} {\bibinfo {author} {\bibfnamefont {J.}~\bibnamefont
  {Bostwick}}\ and\ \bibinfo {author} {\bibfnamefont {P.}~\bibnamefont
  {Steen}},\ }\href@noop {} {\bibfield  {journal} {\bibinfo  {journal} {J.
  Fluid Mech.}\ }\textbf {\bibinfo {volume} {760}},\ \bibinfo {pages} {5}
  (\bibinfo {year} {2014})}\BibitemShut {NoStop}%
\bibitem [{\citenamefont {Bostwick}\ and\ \citenamefont
  {Steen}(2015)}]{bostwickARFM}%
  \BibitemOpen
  \bibfield  {author} {\bibinfo {author} {\bibfnamefont {J.}~\bibnamefont
  {Bostwick}}\ and\ \bibinfo {author} {\bibfnamefont {P.}~\bibnamefont
  {Steen}},\ }\href@noop {} {\bibfield  {journal} {\bibinfo  {journal} {Ann.
  Rev. Fluid Mech.}\ }\textbf {\bibinfo {volume} {47}},\ \bibinfo {pages} {539}
  (\bibinfo {year} {2015})}\BibitemShut {NoStop}%
\bibitem [{\citenamefont {Tanner}(1979)}]{tanner79}%
  \BibitemOpen
  \bibfield  {author} {\bibinfo {author} {\bibfnamefont {L.}~\bibnamefont
  {Tanner}},\ }\href@noop {} {\bibfield  {journal} {\bibinfo  {journal} {J.
  Phys. D: Appl. Phys.}\ }\textbf {\bibinfo {volume} {12}},\ \bibinfo {pages}
  {1473} (\bibinfo {year} {1979})}\BibitemShut {NoStop}%
\bibitem [{\citenamefont {Chen}(1988)}]{chen88}%
  \BibitemOpen
  \bibfield  {author} {\bibinfo {author} {\bibfnamefont {C.}~\bibnamefont
  {Chen}},\ }\href@noop {} {\bibfield  {journal} {\bibinfo  {journal} {Journal
  of Colloid and Interface Science}\ }\textbf {\bibinfo {volume} {122}},\
  \bibinfo {pages} {60} (\bibinfo {year} {1988})}\BibitemShut {NoStop}%
\bibitem [{\citenamefont {Ehrhard}\ and\ \citenamefont
  {Davis}(1991)}]{ehrhard91}%
  \BibitemOpen
  \bibfield  {author} {\bibinfo {author} {\bibfnamefont {P.}~\bibnamefont
  {Ehrhard}}\ and\ \bibinfo {author} {\bibfnamefont {S.}~\bibnamefont
  {Davis}},\ }\href@noop {} {\bibfield  {journal} {\bibinfo  {journal} {J.
  Fluid Mech.}\ }\textbf {\bibinfo {volume} {229}},\ \bibinfo {pages} {365}
  (\bibinfo {year} {1991})}\BibitemShut {NoStop}%
\bibitem [{\citenamefont {Bostwick}(2013)}]{bostwick2013thermal}%
  \BibitemOpen
  \bibfield  {author} {\bibinfo {author} {\bibfnamefont {J.}~\bibnamefont
  {Bostwick}},\ }\href@noop {} {\bibfield  {journal} {\bibinfo  {journal}
  {Journal of Fluid Mechanics}\ }\textbf {\bibinfo {volume} {725}},\ \bibinfo
  {pages} {566} (\bibinfo {year} {2013})}\BibitemShut {NoStop}%
\bibitem [{\citenamefont {Hocking}(1987{\natexlab{a}})}]{hocking1987damping}%
  \BibitemOpen
  \bibfield  {author} {\bibinfo {author} {\bibfnamefont {L.}~\bibnamefont
  {Hocking}},\ }\href@noop {} {\bibfield  {journal} {\bibinfo  {journal}
  {Journal of fluid mechanics}\ }\textbf {\bibinfo {volume} {179}},\ \bibinfo
  {pages} {253} (\bibinfo {year} {1987}{\natexlab{a}})}\BibitemShut {NoStop}%
\bibitem [{\citenamefont {Emslie}\ \emph {et~al.}(1958)\citenamefont {Emslie},
  \citenamefont {Bonner},\ and\ \citenamefont {Peck}}]{emslie58}%
  \BibitemOpen
  \bibfield  {author} {\bibinfo {author} {\bibfnamefont {A.}~\bibnamefont
  {Emslie}}, \bibinfo {author} {\bibfnamefont {F.}~\bibnamefont {Bonner}}, \
  and\ \bibinfo {author} {\bibfnamefont {L.}~\bibnamefont {Peck}},\ }\href@noop
  {} {\bibfield  {journal} {\bibinfo  {journal} {J. Applied Physics}\ }\textbf
  {\bibinfo {volume} {29}},\ \bibinfo {pages} {858} (\bibinfo {year}
  {1958})}\BibitemShut {NoStop}%
\bibitem [{\citenamefont {Schwartz}\ and\ \citenamefont
  {Roy}(2004)}]{schwartz}%
  \BibitemOpen
  \bibfield  {author} {\bibinfo {author} {\bibfnamefont {L.}~\bibnamefont
  {Schwartz}}\ and\ \bibinfo {author} {\bibfnamefont {R.}~\bibnamefont {Roy}},\
  }\href@noop {} {\bibfield  {journal} {\bibinfo  {journal} {Physics of
  Fluids}\ }\textbf {\bibinfo {volume} {16}},\ \bibinfo {pages} {569} (\bibinfo
  {year} {2004})}\BibitemShut {NoStop}%
\bibitem [{\citenamefont {Froehlich}(2009)}]{froehlich2009two}%
  \BibitemOpen
  \bibfield  {author} {\bibinfo {author} {\bibfnamefont {M.}~\bibnamefont
  {Froehlich}},\ }\emph {\bibinfo {title} {Two coating problems: Thin film
  rupture and spin coating}},\ \href@noop {} {Ph.D. thesis},\ \bibinfo
  {school} {Duke University} (\bibinfo {year} {2009})\BibitemShut {NoStop}%
\bibitem [{\citenamefont {Melo}\ \emph {et~al.}(1989)\citenamefont {Melo},
  \citenamefont {Joanny},\ and\ \citenamefont {Fauvre}}]{melo}%
  \BibitemOpen
  \bibfield  {author} {\bibinfo {author} {\bibfnamefont {F.}~\bibnamefont
  {Melo}}, \bibinfo {author} {\bibfnamefont {J.}~\bibnamefont {Joanny}}, \ and\
  \bibinfo {author} {\bibfnamefont {S.}~\bibnamefont {Fauvre}},\ }\href@noop {}
  {\bibfield  {journal} {\bibinfo  {journal} {Physical Review Letters}\
  }\textbf {\bibinfo {volume} {63}},\ \bibinfo {pages} {1958} (\bibinfo {year}
  {1989})}\BibitemShut {NoStop}%
\bibitem [{\citenamefont {Fraysse}\ and\ \citenamefont
  {Homsy}(1994)}]{fraysse94}%
  \BibitemOpen
  \bibfield  {author} {\bibinfo {author} {\bibfnamefont {N.}~\bibnamefont
  {Fraysse}}\ and\ \bibinfo {author} {\bibfnamefont {G.}~\bibnamefont
  {Homsy}},\ }\href@noop {} {\bibfield  {journal} {\bibinfo  {journal} {Phys.
  Fluids}\ }\textbf {\bibinfo {volume} {6}},\ \bibinfo {pages} {1491} (\bibinfo
  {year} {1994})}\BibitemShut {NoStop}%
\bibitem [{\citenamefont {Spaid}\ and\ \citenamefont {Homsy}(1996)}]{spaid96}%
  \BibitemOpen
  \bibfield  {author} {\bibinfo {author} {\bibfnamefont {M.}~\bibnamefont
  {Spaid}}\ and\ \bibinfo {author} {\bibfnamefont {G.}~\bibnamefont {Homsy}},\
  }\href@noop {} {\bibfield  {journal} {\bibinfo  {journal} {Phys. Fluids}\
  }\textbf {\bibinfo {volume} {9}},\ \bibinfo {pages} {823} (\bibinfo {year}
  {1996})}\BibitemShut {NoStop}%
\bibitem [{\citenamefont {McKinley}\ and\ \citenamefont
  {Wilson}(2002)}]{mckinley2002linear}%
  \BibitemOpen
  \bibfield  {author} {\bibinfo {author} {\bibfnamefont {I.}~\bibnamefont
  {McKinley}}\ and\ \bibinfo {author} {\bibfnamefont {S.}~\bibnamefont
  {Wilson}},\ }\href@noop {} {\bibfield  {journal} {\bibinfo  {journal}
  {Physics of Fluids (1994-present)}\ }\textbf {\bibinfo {volume} {14}},\
  \bibinfo {pages} {133} (\bibinfo {year} {2002})}\BibitemShut {NoStop}%
\bibitem [{\citenamefont {Boettcher}\ and\ \citenamefont
  {Ehrhard}(2014)}]{boettcher2014contact}%
  \BibitemOpen
  \bibfield  {author} {\bibinfo {author} {\bibfnamefont {K.~E.}\ \bibnamefont
  {Boettcher}}\ and\ \bibinfo {author} {\bibfnamefont {P.}~\bibnamefont
  {Ehrhard}},\ }\href@noop {} {\bibfield  {journal} {\bibinfo  {journal}
  {European Journal of Mechanics-B/Fluids}\ }\textbf {\bibinfo {volume} {43}},\
  \bibinfo {pages} {33} (\bibinfo {year} {2014})}\BibitemShut {NoStop}%
\bibitem [{\citenamefont {Dussan~V.}\ and\ \citenamefont
  {Davis}(1974)}]{dussan74}%
  \BibitemOpen
  \bibfield  {author} {\bibinfo {author} {\bibfnamefont {E.}~\bibnamefont
  {Dussan~V.}}\ and\ \bibinfo {author} {\bibfnamefont {S.}~\bibnamefont
  {Davis}},\ }\href@noop {} {\bibfield  {journal} {\bibinfo  {journal} {J.
  Fluid Mech.}\ }\textbf {\bibinfo {volume} {65}},\ \bibinfo {pages} {71}
  (\bibinfo {year} {1974})}\BibitemShut {NoStop}%
\bibitem [{\citenamefont {Popescu}\ \emph {et~al.}(2012)\citenamefont
  {Popescu}, \citenamefont {Oshanin}, \citenamefont {Dietrich},\ and\
  \citenamefont {Cazabat}}]{popescu2012precursor}%
  \BibitemOpen
  \bibfield  {author} {\bibinfo {author} {\bibfnamefont {M.~N.}\ \bibnamefont
  {Popescu}}, \bibinfo {author} {\bibfnamefont {G.}~\bibnamefont {Oshanin}},
  \bibinfo {author} {\bibfnamefont {S.}~\bibnamefont {Dietrich}}, \ and\
  \bibinfo {author} {\bibfnamefont {A.}~\bibnamefont {Cazabat}},\ }\href@noop
  {} {\bibfield  {journal} {\bibinfo  {journal} {Journal of Physics: Condensed
  Matter}\ }\textbf {\bibinfo {volume} {24}},\ \bibinfo {pages} {243102}
  (\bibinfo {year} {2012})}\BibitemShut {NoStop}%
\bibitem [{\citenamefont {Hocking}(1987{\natexlab{b}})}]{hocking1987waves}%
  \BibitemOpen
  \bibfield  {author} {\bibinfo {author} {\bibfnamefont {L.}~\bibnamefont
  {Hocking}},\ }\href@noop {} {\bibfield  {journal} {\bibinfo  {journal}
  {Journal of fluid mechanics}\ }\textbf {\bibinfo {volume} {179}},\ \bibinfo
  {pages} {267} (\bibinfo {year} {1987}{\natexlab{b}})}\BibitemShut {NoStop}%
\bibitem [{\citenamefont {Rosenblat}\ and\ \citenamefont
  {Davis}(1985)}]{rosenblat}%
  \BibitemOpen
  \bibfield  {author} {\bibinfo {author} {\bibfnamefont {S.}~\bibnamefont
  {Rosenblat}}\ and\ \bibinfo {author} {\bibfnamefont {S.}~\bibnamefont
  {Davis}},\ }in\ \href@noop {} {\emph {\bibinfo {booktitle} {Frontiers in
  Fluid Mechanics}}},\ \bibinfo {editor} {edited by\ \bibinfo {editor}
  {\bibfnamefont {S.}~\bibnamefont {Davis}}\ and\ \bibinfo {editor}
  {\bibfnamefont {J.}~\bibnamefont {Lumley}}}\ (\bibinfo  {publisher} {Springer
  Verlag},\ \bibinfo {year} {1985})\ pp.\ \bibinfo {pages}
  {171--183}\BibitemShut {NoStop}%
\bibitem [{\citenamefont {Smith}(1995)}]{smith95}%
  \BibitemOpen
  \bibfield  {author} {\bibinfo {author} {\bibfnamefont {M.}~\bibnamefont
  {Smith}},\ }\href@noop {} {\bibfield  {journal} {\bibinfo  {journal} {J.
  Fluid Mech.}\ }\textbf {\bibinfo {volume} {294}},\ \bibinfo {pages} {209}
  (\bibinfo {year} {1995})}\BibitemShut {NoStop}%
\bibitem [{\citenamefont {Lacey}\ \emph {et~al.}(1982)\citenamefont {Lacey},
  \citenamefont {Ockendon},\ and\ \citenamefont {Tayler}}]{lacey1982waiting}%
  \BibitemOpen
  \bibfield  {author} {\bibinfo {author} {\bibfnamefont {A.}~\bibnamefont
  {Lacey}}, \bibinfo {author} {\bibfnamefont {J.}~\bibnamefont {Ockendon}}, \
  and\ \bibinfo {author} {\bibfnamefont {A.}~\bibnamefont {Tayler}},\
  }\href@noop {} {\bibfield  {journal} {\bibinfo  {journal} {SIAM Journal on
  Applied Mathematics}\ }\textbf {\bibinfo {volume} {42}},\ \bibinfo {pages}
  {1252} (\bibinfo {year} {1982})}\BibitemShut {NoStop}%
\bibitem [{\citenamefont {Marino}\ \emph {et~al.}(1996)\citenamefont {Marino},
  \citenamefont {Thomas}, \citenamefont {Gratton}, \citenamefont {Diez},
  \citenamefont {Betel{\'u}},\ and\ \citenamefont
  {Gratton}}]{marino1996waiting}%
  \BibitemOpen
  \bibfield  {author} {\bibinfo {author} {\bibfnamefont {B.}~\bibnamefont
  {Marino}}, \bibinfo {author} {\bibfnamefont {L.}~\bibnamefont {Thomas}},
  \bibinfo {author} {\bibfnamefont {R.}~\bibnamefont {Gratton}}, \bibinfo
  {author} {\bibfnamefont {J.}~\bibnamefont {Diez}}, \bibinfo {author}
  {\bibfnamefont {S.}~\bibnamefont {Betel{\'u}}}, \ and\ \bibinfo {author}
  {\bibfnamefont {J.}~\bibnamefont {Gratton}},\ }\href@noop {} {\bibfield
  {journal} {\bibinfo  {journal} {Physical Review E}\ }\textbf {\bibinfo
  {volume} {54}},\ \bibinfo {pages} {2628} (\bibinfo {year}
  {1996})}\BibitemShut {NoStop}%
\bibitem [{\citenamefont {Mukhopadhyay}\ \emph {et~al.}(2011)\citenamefont
  {Mukhopadhyay}, \citenamefont {Murisic}, \citenamefont {Behringer},\ and\
  \citenamefont {Kondic}}]{mukho11}%
  \BibitemOpen
  \bibfield  {author} {\bibinfo {author} {\bibfnamefont {S.}~\bibnamefont
  {Mukhopadhyay}}, \bibinfo {author} {\bibfnamefont {N.}~\bibnamefont
  {Murisic}}, \bibinfo {author} {\bibfnamefont {R.}~\bibnamefont {Behringer}},
  \ and\ \bibinfo {author} {\bibfnamefont {L.}~\bibnamefont {Kondic}},\
  }\href@noop {} {\bibfield  {journal} {\bibinfo  {journal} {Physical Review
  E}\ }\textbf {\bibinfo {volume} {83}},\ \bibinfo {pages} {046302} (\bibinfo
  {year} {2011})}\BibitemShut {NoStop}%
\end{thebibliography}%

\end{document}